\documentclass[pra,twocolumn,showpacs,nofootinbib]{revtex4}

\newcommand{\BR}{\nonumber\\&&}

\usepackage{graphicx}%
\usepackage{dcolumn}
\usepackage{amsmath}
\usepackage{times}

\begin{document}
\bibliographystyle{apsrev}
\preprint{BLAKIE1}

\title[Bragg Spectroscopy]{Theory of coherent Bragg spectroscopy of a
trapped
Bose-Einstein condensate}

\author{P. B. Blakie}
 \email{bblakie@physics.otago.ac.nz}
\affiliation{Department of Physics, University of Otago, Dunedin, New
Zealand}
\author{R. J. Ballagh}
\affiliation{Department of Physics, University of Otago, Dunedin, New
Zealand}

\author{C. W. Gardiner}
\affiliation{School of Chemical and Physical Sciences, Victoria University,
Wellington, New Zealand}
\date{\today}

\begin{abstract}

We present a detailed theoretical analysis of Bragg spectroscopy
from a Bose-Einstein condensate at $T=0$K. We demonstrate that
within the linear response regime, both a quantum field theory
treatment and a meanfield Gross-Pitaevskii treatment lead to the
same value for the mean evolution of the quasiparticle operators.
The observable for Bragg spectroscopy experiments, which is the
spectral response function of the momentum transferred to the
condensate, can therefore be calculated in a meanfield formalism.
We analyse the behaviour of this observable by carrying out
numerical simulations in axially symmetric three-dimensional cases
and in two dimensions.  An approximate analytic expression for the
observable is obtained and provides a means for identifying the
relative importance of three broadening and shift mechanisms
(meanfield, Doppler, and finite pulse duration) in different
regimes.  We show that the suppression of scattering at small
values of $q$ observed by Stamper-Kurn {\it et~al.}, [Phys.  Rev.
Lett.  {\bf 83}, 2876 (1999)] is accounted for by the meanfield
treatment, and can be interpreted in terms of the interference of
the $u$ and $v$ quasiparticle amplitudes.  We also show that,
contrary to the assumptions of previous analyses, there is no
regime for trapped condensates for which the spectral response
function and the dynamic structure factor are equivalent.  Our
numerical calculations can also be performed outside the linear
response regime, and show that at large laser intensities a
significant decrease in the shift of the spectral response
function can occur due to depletion of the initial condensate.

\end{abstract}

\maketitle

\section{Introduction}

In 1999 Ketterle's group at MIT reported a set of experiments in
which condensate properties were measured using the technique of
Bragg spectroscopy \cite{Strenger99, Stamper99}. In those
experiments a low intensity Bragg pulse was used to excite a small
amount of condensate into a higher momentum state, and the
\emph{Bragg spectrum} of the condensate was found by measuring the
momentum transfer for a range of Bragg frequencies ($\omega $) and
momenta ($\hbar \mathbf{q}$). That work established Bragg
spectroscopy as a tool capable of measuring condensate properties
with spectroscopic precision. The theoretical analysis of the
measurements however, gives rise to a number of issues. Ketterle
and his colleagues assumed that the spectra gave a direct
measurement of the \emph{dynamic structure factor}, which is the
Fourier transform of the density-density correlation function, and
is familiar as the observable in neutron scattering experiments in
super-fluid helium\cite{Pines69,Pines69b,Griffin93}. They also
attributed the suppression of imparted momentum they observed at
low $q$ values to correlated pair excitations, and quantum
depletion of the condensate, and speculated \cite{Ketterle00} that
an accurate description would require a more complete quantum
treatment. The purpose of the current paper is to develop a theory
of Bragg spectroscopy that is valid in the regime of the
experiments, and to use it to make quantitative calculations and
to analyze the phenomena that occur in this regime. We also
investigate the relationship between the observable of Bragg
spectroscopy (i.e. the momentum transferred to the condensate) and
the dynamic structure factor, and show that for a {\em trapped}
condensate there is no regime in which one can simply be obtained
from the other.

We begin in section \ref{SEC_THEORY} within the framework of many
body field theory and calculate, in the Bogoliubov approximation,
the linear response of the condensate to an applied Bragg pulse.
We obtain expressions for the temporal evolution of the
quasiparticle operators, and show that they have a nonzero mean
value, i.e. that the quasiparticles are generated as coherent
states. We demonstrate that within a well defined regime the mean
values of the Bogoliubov operators are identical to the amplitudes
obtained from a linearized mean-field (Gross-Pitaevskii)
treatment. The meanfield treatment therefore provides a valid
description of the experiments in the regime of small excitation.

A number of meanfield theoretical treatments of Bragg scattering
from condensates have been given. Blakie and Ballagh have
presented a quantitative meanfield description \cite{Blakie00}
which confirmed the analysis of the Bragg spectroscopy shift given
by the MIT group, and provided analytic estimates for a number of
quantities, including the momentum width of the scattered
condensate. Stringari and colleagues have also used a meanfield
description to analyze Bragg spectroscopy
\cite{Zambelli00,Brunello2001}, and in addition have used the
approach to devise schemes for measuring quasiparticle amplitudes
\cite{Brunello00}, and for making spatially separate condensates
interfere \cite{Pitaevskii1999b}. In the current paper we use the
Gross Pitaevskii formulation of Bragg scattering as presented by
Blakie and Ballagh \cite{Blakie00} to analyze the behavior
observed in the Bragg spectroscopy experiments.

The observable in the experiments is the momentum
transferred to the condensate, and in section \ref{SECmeasurecond} we define
a normalized version of the expectation value of this quantity
that we call the \emph{spectral response function} $R(\mathbf{q},\omega )$.
For a trapped condensate the momentum
transfer can arise from two sources, the Bragg beams or the trap itself,
which complicates the analysis. The MIT group recognized this issue, and
applied the Bragg laser pulses only for a small fraction of the trap period,
and then released the trap. However the tradeoff involved in minimizing
momentum transfer from the trap by using a short Bragg pulse significantly
compromises the energy selectivity of the process. We examine the influence
this has on the relationship between $R(\mathbf{q},\omega )$ and the dynamic
structure factor $S(\mathbf{q},\omega )$, and we show that in the presence
of a trap, the evaluation of $S(\mathbf{q},\omega )$ requires
$R(\mathbf{q},\omega )$ to be known for all possible pulse lengths.

The central quantity of Bragg spectroscopy is thus the spectral
response function $R(\mathbf{q},\omega )$ and we derive an
approximate analytic expression for this quantity, incorporating
the effects of both the meanfield interaction and the finite
duration of the Bragg pulse, in section \ref{SEC_QH}. In section
\ref{SECBraggspec} we use $R(\mathbf{q},\omega )$ to characterize
our numerical investigations of Bragg spectroscopy, and we
consider a wide range of three dimensional axially symmetric
scenarios, for which we simulate the experiments using the
meanfield (Gross Pitaevskii) equation for Bragg scattering
\cite{Blakie00}. We verify the validity range of our approximate
form for $R(\mathbf{q},\omega )$ by comparing it to the full
numerical results, and we identify the regimes in which one or
other of the mechanisms of: the meanfield interaction, the Doppler
effect, and the finite pulse duration, dominates the formation of
the Bragg spectrum. We also show that our approximate form for
$R(\mathbf{q},\omega )$ will allow a more accurate estimation of
the momentum width of a condensate than obtained by previous
analyses.

Our numerical simulations allow us to calculate the effect on Bragg
spectroscopy of laser intensities sufficiently large that linear response
theory no longer holds. We investigate cases where the scattered fraction of
the condensate is of order 20\%, and show that the depletion of the ground
state condensate leads to a significant reduction of the frequency
shift, which has not been accounted for in previous analyses.  We also
consider the spectral response function from a vortex, using two
dimensional simulations.  Finally in section \ref{SEC_long_time} we
investigate the energy response of a condensate subject to a Bragg
pulse of sufficiently long duration that individual quasiparticle
excitations can be resolved.

\section{Low-intensity Bragg scattering theory}\label{SEC_THEORY}

In this section we calculate the response of the condensate to a Bragg
pulse within the linear regime, using two distinct approaches.  In the
first of these (section \ref{SEC_THEORY}A) we use the many body field
theory formalism in the Bogoliubov approximation, to calculate the temporal
evolution of the quasiparticle operators.  In the second approach (section
\ref{SEC_THEORY}B) we
use a meanfield (Gross Pitaevskii) equation and obtain the amplitudes of
the linearized response.  The two approaches are shown to give identical
mean results in section \ref{SEC_THEORY}C.

\subsection{Many-body field theoretic approach}
\begin{figure}
{\centering \includegraphics{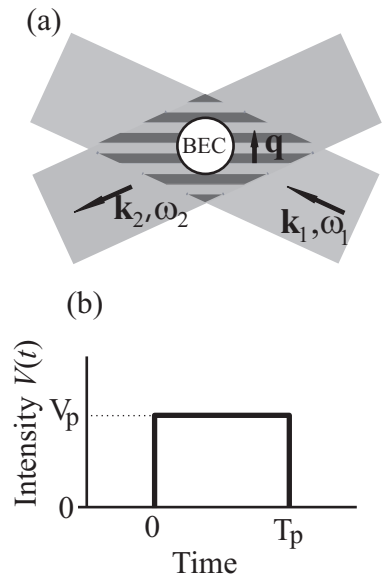}
\par}
\caption{\label{ExptDiagram} Bragg spectroscopy of a Bose-Einstein
condensate (BEC). (a) Two laser beams with wavevectors
$\mathbf{k}_1$ and $\mathbf{k}_2$ and frequencies $\omega_1$ and
$\omega_2$ respectively, create a moving optical potential
 with
wavevector $\mathbf{q}=\mathbf{k}_1-\mathbf{k}_2$ and frequency $
\omega=\omega_1-\omega_2$ (see \cite{Blakie00}). (b) Temporal
behaviour of the Bragg pulse assumed in this paper. }
\end{figure}

The many-body Hamiltonian for $N_{0}$ identical bosons in a trap and subject
to a time-dependent Bragg pulse can be written%
\begin{equation}
\hat{H}=\hat{H}_{0}+\hat{H}_{I}(t),  \label{EQN_H_equals_H0_plus_HI}
\end{equation}%
where $\hat{H}_{0}$ is the usual trapped boson Hamiltonian%
\begin{equation}
\hat{H}_{0}=\int d\mathbf{r}\,\hat{\Psi}^{\dagger }\left[ -\frac{\hbar ^{2}}{%
2m}\nabla ^{2}+V_{T}(\mathbf{r})\right] \hat{\Psi}+\frac{U_{0}}{2}\int d%
\mathbf{r}\,\hat{\Psi}^{\dagger }\hat{\Psi}^{\dagger }\hat{\Psi}\hat{\Psi}.
\label{EQN_MB_H0}
\end{equation}%
$V_{T}(\mathbf{r})$ is the trapping potential, which we choose to
be harmonic. The Bragg interaction $\hat{H}_{I}(t)$ arises from
two overlapping plane wave laser fields, which have equal
amplitudes but frequency and wave vector differences $\omega $ and
$\mathbf{q}$ respectively (see Fig. \ref{ExptDiagram}(a)). The
laser fields are treated classically, and their interaction with
the internal transition of the atoms is characterized by a Rabi
frequency $\Omega (t)$ (for the combined fields at the intensity
peaks) and a detuning $\Delta $ which is large and essentially the
same for both laser fields. In this regime the internal structure
for the atoms can be eliminated (see \cite{Blakie00} for details)
so that the field operator $\hat{\Psi}$ refers only to the ground
internal state, and $\hat{H}_{I}(t)$ takes the form
\begin{equation}
\hat{H}_{I}(t)=\int d\mathbf{r}\,\hat{\Psi}^{\dagger }\left[ \hbar V(t)\cos (%
\mathbf{q}\cdot \mathbf{r}-\omega t)\right] \hat{\Psi},  \label{EQN_MB_HB}
\end{equation}%
where $V(t)=\Omega ^{2}(t)/2|\Delta |$ (see Fig.
\ref{ExptDiagram}(a)).

\subsubsection{Bogoliubov transformation}\label{SEC_BOG_TRANSFORMATION}

For a highly occupied stationary state (\( T\approx 0\)K) the
field operator can be written in the Bogoliubov approximation as a
sum of meanfield and operator parts (\( \hat{\Psi }=\langle
\hat{\Psi }\rangle +\hat{\phi } \)). In this paper we mostly
consider a ground state, but we also consider the case of a
central vortex. Following standard treatments (e.g.
\cite{Griffin96, Fetter72}) we employ a Bogoliubov transformation
for the operator part to write
\begin{eqnarray}
\label{EQN_BOG_EXPN}
\hat{\Psi }(\mathbf{r},t)&=&\sqrt{N_{0}}\psi _{0}(\mathbf{r})e^{-i\mu
t}+e^{i(S_{0}(\mathbf{r})-\mu t)}\times \\
&&\sum _{i}\left(
\hat{b}_{i}(t)e^{-i\omega
_{i}t}\tilde{u}_{i}(\mathbf{r})+\hat{b}^{\dagger }_{i}(t)e^{i\omega
_{i}t}\tilde{v}^{*}_{i}(\mathbf{r})\right) \nonumber,
\end{eqnarray}
 where the condensate is represented by the first term, and
 \( \hat{b}_{i} \) and \( \hat{b}^{\dagger }_{i} \) are the
quasiparticle destruction and creation operators respectively (in
an interaction picture with respect to \( \hat{H}_{0} \)). The
state \( \psi _{0}=|\psi _{0}|\exp (iS_{0}) \), is an eigenstate solution,
with eigenvalue $\hbar\mu$ of the time independent Gross-Pitaevskii equation
\begin{equation}
\label{EQN_TIGPE}
\hbar \mu
\psi_{0}=\left[-\frac{\hbar^2}{2m}\nabla^2+V_T(\mathbf{r})\right]\psi_0+
N_0U_0|\psi_0|^2\psi_0,
\end{equation}
and the functions \(\{\tilde{u}_{i},\tilde{v}_{i}\} \)
are the \emph{orthogonal} quasiparticle basis states \cite{Morgan98}.
These basis states are orthogonal to the condensate mode and are related
to the usual (non-orthogonal) quasiparticle basis states \( \{u_{i},v_{i}\} \)
(given below) by projection into the subspace orthogonal to the
condensate,
i.e. \begin{eqnarray}
\tilde{u}_{i} & = & u_{i}-a_{i}|\psi
_{0}|,\label{EQN_Ortho_QP1} \\
\tilde{v}_{i}^{*} & = & v_{i}^{*}+a^{*}_{i}|\psi
_{0}|,\label{EQN_Ortho_QP2}
\end{eqnarray}
where \begin{equation}
\label{EQN_aj}
a_{i}=\int d\mathbf{r}\, |\psi _{0}|u_{i}=-\int d\mathbf{r}\, |\psi
_{0}|v_{i},
\end{equation}
(see \cite{Morgan98}).

The form of Bogoliubov transformation used in Eq. (\ref{EQN_BOG_EXPN})
 explicitly includes the phase (\( S_{0}\))
of \( \psi _{0} \), which is convenient in cases where \( \psi _{0} \)
is not necessarily a ground state. We note that a number of different sign
conventions appear in the literature, and ours differs from that in Ref.
\cite{Fetter72}.
We discuss the different conventions in the Appendix.
 The Bogoliubov-de Gennes equation
for \( \{u_{i},v_{i}\} \) are \begin{eqnarray}
{\cal L}u_{i}+N_{0}U_{0}|\psi _{0}|^{2}v_{i} & = & \hbar \omega
_{i}u_{i},\label{EQN_QP_BdeG_EQN1} \\
{\cal L}^{*}v_{i}+N_{0}U_{0}|\psi _{0}|^{2}u_{i} & = & -\hbar \omega
_{i}v_{i},\label{EQN_QP_BdeG_EQN2}
\end{eqnarray}
where\begin{equation}
\label{EQN_calL}
{\cal L}=\left[ -{\hbar ^{2}\over 2m}(\nabla +i\nabla
S_{0})^{2}+V_{T}(\mathbf{r})-\hbar \mu +2N_{0}U_{0}|\psi
_{0}|^{2}\right] ,
\end{equation}
 with the orthogonality conditions\begin{eqnarray}
\int d\mathbf{r}\{u_{i}u_{j}^{*}-v_{i}v_{j}^{*}\} & = & \delta
_{ij},\label{EQN_orthog_uv_cond1} \\
\int d\mathbf{r}\{u_{i}v_{j}-v_{i}u_{j}\} & = &
0.\label{EQN_orthog_uv_cond2}
\end{eqnarray}

\subsubsection{Bogoliubov Hamiltonian}

Applying the transformation in Eq. (\ref{EQN_BOG_EXPN}), to
the Hamiltonians (\ref{EQN_MB_H0}) and (\ref{EQN_MB_HB})
we obtain their Bogoliubov form, i.e.
\begin{eqnarray}
\hat{H}_{0}&\approx& \hat{H}_{0}^{B} \nonumber\\& \equiv  & E_{0}+\sum _{i}\hbar
\omega _{i}\hat{b}_{i}^{\dagger
}\hat{b}_{i},\label{EQN_H0_BOG} \\
\hat{H}_{I}(t)&\approx& \hat{H}^{B}_{I}(t)\nonumber\\ & \equiv  & N_{0}\int
d\mathbf{r}\, |\psi _{0}|^{2}\hbar V(t)\cos (\mathbf{q}\cdot
\mathbf{r}-\omega t)\label{EQN_HI_BOG} \\
 &  & +\sqrt{N_{0}}\sum _{i}\Big[ \hat{b}_{i}^{\dagger }e^{i\omega
_{i}t}\int d\mathbf{r}\,(\tilde{u}_{i}^{*}+\tilde{v}_{i}^{*})\nonumber \\
&&\times\hbar
V(t)\cos (\mathbf{q}\cdot \mathbf{r}-\omega t)|\psi _{0}|+\rm
{h.c}\Big],\nonumber
\end{eqnarray}
where $E_0$ is the energy of the highly occupied state (see Eq.
(\ref{EQN_energy_sum})). The quasiparticle transformation
diagonalizes \( \hat{H}_{0} \) to quadratic order, and we note
that the orthogonal basis \( \{\tilde{u}_{i},\tilde{v}_{i}\} \) is
required for this diagonalization to be valid. In evaluating \(
\hat{H}^{B}_{I} \), terms involving products of quasiparticle
operators have been ignored. This amounts to neglecting Bragg
induced scattering between quasiparticle states, which is of order
\( 1/\sqrt{N_{0}} \) smaller than the terms linear in \(
\hat{b}_{i}^{\dagger } \) (or \( \hat{b}_{i} \)). Those linear
terms are of primary interest here, as they describe the
scattering between the condensate and quasiparticle states which
occurs as a result of the energy and momentum transfer from the
optical potential.

The time dependent exponentials, \( \exp (\pm i\omega _{i}t) \),
which multiply the quasiparticle operators in Eq.
(\ref{EQN_BOG_EXPN})
account for the free evolution due to \( \hat{H}^{B}_{0} \), and
so the Heisenberg equation

\begin{equation}
\label{EQN_Hei1}
i\hbar \frac{\partial }{\partial t}\left( \hat{b}_{i}e^{-i\omega
_{i}t}\right) =\left[ \left( \hat{b}_{i}e^{-i\omega _{i}t}\right)
,\hat{H}_{0}^{B}+\hat{H}^{B}_{I}(t)\right] ,
\end{equation}
becomes\begin{eqnarray}
i\hbar \frac{\partial }{\partial t}\hat{b}_{i} & = & \left[
\hat{b}_{i},\hat{H}_{0}^{B}+\hat{H}^{B}_{I}(t)\right] -\hbar \omega
_{i}\hat{b}_{i},\\
 & = & \left[ \hat{b}_{i},\hat{H}_{I}^{B}(t)\right]
.\label{EQN_Hei_EQN_of_Motion}
\end{eqnarray}
This is easily solved to give

\begin{equation}
\label{EQN_bi_from_Heisen_Eq_of_Mot}
\hat{b}_{i}(t)=\hat{b}_{i}(0)-\beta _{i}(t),
\end{equation}
 where \( \beta _{i}(t) \) is a \( c \)-number,

\begin{eqnarray}
\label{EQN_beta_coherent_amp}
\beta _{i}(t)&=&-i\sqrt{N_{0}}\int _{0}^{t}dt^{\prime }\, V(t^{\prime
})e^{i\omega _{i}t^{\prime }}\int d\mathbf{r}\, \cr
&& \times(\tilde{u}^{*}_{i}+\tilde{v}^{*}_{i})\cos (\mathbf{q}\cdot
\mathbf{r}-\omega t^{'})|\psi _{0}|.
\end{eqnarray}
We see that the Bragg excitation causes the quasiparticle operators
to develop nonzero mean values. Note that this is a complete solution of the
physics in the linearized regime, from which any observable quantities
can be computed.

\subsubsection{Initial conditions }

Our derivation so far has been based on a \( T=0 \)K Bogoliubov
treatment.
For this case the initial state is the quasiparticle vacuum state (\(
|0\rangle  \)).
Had we started with this initial condition and considered evolution
in the Schr\"odinger picture we would have found that the system
evolves
as \( |0\rangle \rightarrow \exp (i\theta )|\{\beta _{i}\}\rangle  \),
where \( |\{\beta _{i}\}\rangle  \) is a coherent state, i.e. \(
\hat{b}_{i}|\{\beta _{i}\}\rangle =\beta _{i}|\{\beta _{i}\}\rangle
\)
and \( \theta  \) is some phase factor (see \cite{QuantumNoise00}).

Equation (\ref{EQN_bi_from_Heisen_Eq_of_Mot}) also provides
insight into finite temperature cases where most of the atoms are
in the condensate. To fully treat the finite temperature case, it
is necessary to generalize the Bogoliubov treatment to account for
the thermal depletion (e.g. see \cite{Griffin96}), requiring the
functions \( \psi _{0} \), \( u_{i} \), \( v_{i} \) and \( \omega
_{i} \)
to be solved for in a self-consistent manner (e.g. see
\cite{Hutchinson1997a,Hutchinson1998a,Morgan2000}).
In that case, Eq. (\ref{EQN_bi_from_Heisen_Eq_of_Mot}) for
the evolution of \( \hat{b}_{i} \) will still apply, and thus we
see that the initial statistics of \( \hat{b}_{i}(0) \) are preserved
and the mean value \( \langle \hat{b}_{i}\rangle  \) is shifted by
\( \beta _{i} \).

\subsection{Gross-Pitaevskii equation approach}

The Gross-Pitaevskii equation for a condensate of \( N_{0} \)
particles
subject to a Bragg pulse was derived in
\cite{Blakie00}, i.e.
\begin{eqnarray}
\label{EQN_TD_GPE}
i\hbar {\partial \over \partial t}\Psi (\mathbf{r},t)&=&\left[ -{\hbar
^{2}\over 2m}\nabla ^{2}+V_{T}(\mathbf{r})+U_{0}|\Psi |^{2}\right]
\Psi (\mathbf{r},t)\nonumber\\ &&+\hbar V(t)\cos (\mathbf{q}\cdot \mathbf{r}-\omega
t)\Psi (\mathbf{r},t),
\end{eqnarray}
 where \( \Psi (\mathbf{r},t) \) is the condensate meanfield
wavefunction for the atoms in their internal ground state, and is
normalized according to \( \int d\mathbf{r}\, |\Psi |^{2}=N_{0} \).
The condensate wavefunction can be expanded in terms of a
quasiparticle
basis in the form\begin{eqnarray}
\label{EQN_GPE_BOG_DECOMPOSITION}
\Psi (\mathbf{r},t)&=&\sqrt{N_{0}}\psi _{0}e^{-i\mu
t}+e^{i(S_{0}(\mathbf{r})-\mu t)}\\
&&\times\sum _{i}\left(
c_{i}(t)u_{i}e^{-i\omega _{i}t}+c^{*}_{i}(t)v_{i}^{*}e^{i\omega
_{i}t}\right),\nonumber
\end{eqnarray}
where \( c_{i} \) are the time dependent quasiparticle amplitudes.
This expansion has been made using the non-orthogonal quasiparticle
basis, and the ground state has been assumed static. This latter
assumption
will only be valid while excitation induced by the Bragg pulse remains
small. The wavefunction decomposition (Eq.
(\ref{EQN_GPE_BOG_DECOMPOSITION}))
transforms the Gross-Pitaevskii Eq. (\ref{EQN_TD_GPE}) into
\begin{eqnarray}
\label{EQN_GP_IN_QP_Basis}
i\hbar &&\sum _{i}\left( \dot{c}_{i}(t)u_{i}e^{-i\omega
_{i}t}+\dot{c}^{*}_{i}(t)v_{i}^{*}e^{i\omega _{i}t}\right)\\
&&=e^{-i(S_{0}(\mathbf{r})-\mu t)}\hbar V(t)\cos (\mathbf{q}\cdot
\mathbf{r}-\omega t)\Psi (\mathbf{r},t).
\end{eqnarray}
 Morgan \emph{et al.} \cite{Morgan98} have shown how to use the orthogonality
relations
of the quasiparticles, Eqs. (\ref{EQN_orthog_uv_cond1}) and
(\ref{EQN_orthog_uv_cond2}), to project out the quasiparticle
amplitudes from a condensate wavefunction, namely \begin{eqnarray}
\label{EQN_proj_from_Psi}
c_{j}(t)&=&e^{i\omega _{j}t}\int d\mathbf{r}\, \left[ e^{i(\mu
t-S_{0}(\mathbf{r}))}u_{j}^{*}\Psi -v_{j}^{*}e^{-i(\mu
t-S_{0}(\mathbf{r}))}\Psi ^{*}\right] \nonumber\\&&-2a_{j}^{*},
\end{eqnarray}
 where \( a_{j} \) is defined in Eq. (\ref{EQN_aj}). Since
the quasiparticles form a complete set, we may use projection to obtain
a set of equations for quasiparticle amplitudes that are equivalent to
Eq. (\ref{EQN_GP_IN_QP_Basis}), namely
\begin{eqnarray}
\dot{c}_{j}(t) & = & -iV(t)e^{i\omega _{i}t}\int
d\mathbf{r}\, \Big [u_{j}^{*}e^{i(\mu t-S_{0}(\mathbf{r}))}\cos
(\mathbf{q}\cdot \mathbf{r}-\omega t)\Psi \nonumber \\
 &  & +v_{j}^{*}e^{-i(\mu t-S_{0}(\mathbf{r}))}\cos
(\mathbf{q}\cdot \mathbf{r}-\omega t)\Psi  ^{*}\Big
].\label{EQN_GP_proj}
\end{eqnarray}
Because the quasiparticles occupations are all small compared to the
condensate mode, we can simplify Eq. (\ref{EQN_GP_proj}) by setting \( \Psi
(\mathbf{r},t)=\sqrt{N_{0}}|\psi _{0}(\mathbf{r})|\exp
(iS_{0}(\mathbf{r})-i\mu t) \),
yielding \begin{eqnarray}
\label{EQN_ci_from_GP_soln}
c_{i}(t)&=&-i\sqrt{N_{0}}\int _{0}^{t}dt^{\prime }\, V(t^{\prime
})e^{i\omega _{i}t^{\prime }}\int d\mathbf{r}\,
(u^{*}_{i}+v^{*}_{i})\nonumber\\ &&\times \cos (\mathbf{q}\cdot \mathbf{r}-\omega
t^{'})|\psi _{0}|.
\end{eqnarray}
This will of course only provide a good solution while the
quasiparticle
occupations all remain small.

\subsection{Comparison of approaches }

Direct comparison of the Gross-Pitaevskii and quantum field theoretic
results is complicated by the fact that they have been derived for
different basis sets. We note that although the Gross-Pitaevskii
analysis
was carried out using the \( \{u_{i},v_{i}\} \) basis, it is equally
straightforward to use the orthogonal basis \(
\{\tilde{u}_{i},\tilde{v}_{i}\} \).
Morgan \emph{et al.} \cite{Morgan98} have shown that (in analysis
of the Gross-Pitaevskii equation) transforming between these bases
affects only the ground state population and gives rise to no
difference
in the quasiparticle occupations.

Connections between the field theory and simple Gross-Pitaevskii
results based on Eq. (\ref{EQN_TD_GPE}) can only be expected to
exist at \( T=0 \)K, where thermal effects can be ignored
\footnote[1]{We use the word simple here to distinguish our  $ T=0
$K Gross-Pitaevskii equation from more elaborate forms of
Gross-Pitaevskii theory which have been used to investigate finite
temperature effects. For example see \cite{Davis00}. \protect}. In
this regime we begin by considering the relevant vacuum
expectation of the quasiparticle operator from Eq.
(\ref{EQN_bi_from_Heisen_Eq_of_Mot})

\begin{eqnarray}
\label{EQN_bi_expec_on_vac}
\langle \hat{b}_{i}(t)\rangle &=&-i\sqrt{N_{0}}\int _{0}^{t}dt^{\prime
}\, V(t^{\prime })e^{i\omega _{i}t^{\prime }}\int d\mathbf{r}\,
(\tilde{u}^{*}_{i}+\tilde{v}^{*}_{i})\nonumber\\
&&\times\cos (\mathbf{q}\cdot
\mathbf{r}-\omega t^{\prime })|\psi _{0}|.
\end{eqnarray}
This expression emphasizes the coherent nature of the quasiparticle
states, and is the same as the Gross-Pitaevskii result of Eq.
(\ref{EQN_ci_from_GP_soln}),
with the identification \( c_{i}(t)\leftrightarrow \langle
\hat{b}_{i}(t)\rangle  \).
We note that the apparent difference between equations
(\ref{EQN_ci_from_GP_soln})
and (\ref{EQN_bi_expec_on_vac}), where the former depends
on the matrix elements involving \( \{u_{i},v_{i}\} \) and the latter
on matrix elements involving \( \{\tilde{u}_{i},\tilde{v}_{i}\} \),
disappears with the observation that \(
u_{i}+v_{i}=\tilde{u}_{i}+\tilde{v}_{i} \)
(see Eqs. (\ref{EQN_Ortho_QP1}) and
(\ref{EQN_Ortho_QP2})).
Thus we have verified that the \( c_{i}(t) \) and \( \langle
\hat{b}_{i}(t)\rangle  \)
are identical.

\subsection{Quasiparticle occupation}

For the physics we consider in this paper the mean occupation of
the \( i \)-th quasiparticle level (i.e. \( \langle \hat{b}^{\dagger
}_{i}\hat{b}_{i}\rangle  \)
or \( |c_{i}(t)|^{2} \) for the many-body or Gross-Pitaevskii methods
respectively) is of interest, as it is used to make a quasi-homogeneous
approximation to the spectral response function in Eq. (\ref{EQN_homog_mtm_trans}).
Using Eq. (\ref{EQN_bi_from_Heisen_Eq_of_Mot})
we calculate
\begin{eqnarray}
\label{EQN_MBandGP_QP_popn} \langle \hat
n_{i}(t)\rangle&=&N_{0}\Big| \int _{0}^{t}dt^{\prime }\,
e^{i\omega _{i}t^{\prime }}V(t^{\prime })\int d\mathbf{r}\,
(u_{i}^{*}+v_{i}^{*})\nonumber\\
&&\times\cos (\mathbf{q}\cdot \mathbf{r}-\omega
t^{\prime })|\psi _{0}|\Big| ^{2}+\langle \hat n_{i}(0)\rangle ,
\end{eqnarray}
where \( \langle \hat n_{i}(0)\rangle  \) is the initial
occupation, e.g. thermal occupation at \( T\ne 0\)K , and is
included for generality. In Eq. (\ref{EQN_MBandGP_QP_popn}) we
have chosen to use the \( \{u_{i},v_{i}\} \) quasiparticle basis
for ease of comparison with previous results by other authors
(e.g. \cite{Csordas96, Fetter98, Zambelli00, Wu96}).

To progress, it is necessary to specify the temporal behavior of
the Bragg pulse. For simplicity we take the pulse shape as being
square, with  \( V=V_{p} \) for \(0<t<T_p\) (see Fig.
\ref{ExptDiagram}(b)), then Eq. (\ref{EQN_bi_expec_on_vac})
becomes (evaluated at $t=T_p$)
\begin{widetext}
\begin{eqnarray}
\langle \hat{b}_{i}(T_p)\rangle  & = &
-i\sqrt{N_{0}}V_{p}e^{i(\omega _{i}-\omega )T_{p}/2}\Big [\left(
\frac{\sin ((\omega _{i}-\omega )T_{p}/2)}{(\omega _{i}-\omega
)}\right) \label{EQN_bi_vac_expt_finite_Tp} \int d\mathbf{r}\,
(u_{i}^{*}+v_{i}^{*})e^{i\mathbf{q}\cdot \mathbf{r}}|\psi
_{0}|\nonumber \\
 &  & +e^{i\omega T_{p}}\left( \frac{\sin ((\omega _{i}+\omega
)T_{p}/2)}{(\omega _{i}+\omega )}\right) \int d\mathbf{r}\,
(u_{i}^{*}+v_{i}^{*})e^{-i\mathbf{q}\cdot \mathbf{r}}|\psi
_{0}|\Big ].
\end{eqnarray}
\end{widetext}
In this expression, the second term (with denominator \( \omega
_{i}+\omega  \))
will typically be significantly smaller than the first term (with
denominator \( \omega _{i}-\omega  \)) since \( \omega >0 \), so
to a good approximation we can ignore the second term. Similarly,
the quasiparticle occupation result (\ref{EQN_MBandGP_QP_popn})
under the same approximation is
\begin{eqnarray}
\label{EQN_pop_general_square} \langle \hat n_{i}(T_p)\rangle
&=&\frac{\pi V_{p}^{2}T_{p}N_{0}}{2}\left| \int d\mathbf{r}\,
(u_{i}^{*}+v_{i}^{*})e^{i\mathbf{q}\cdot
\mathbf{r}}|\psi _{0}|\right| ^{2} \nonumber\\
&&\times F(\omega _{i}-\omega
,T_{p}) +\langle \hat n_{i}(0)\rangle ,
\end{eqnarray}
 in which\begin{equation}
\label{EQN_time_dep_pert_F}
F(\omega ,T)=\frac{2\sin ^{2}(\omega T/2)}{\pi T\, \omega ^{2}},
\end{equation}
 is a familiar term in time dependent perturbation calculations. In
particular \( F(\omega _{i}-\omega ,T_{p}) \) in Eq.
(\ref{EQN_pop_general_square}) is sharply peaked in frequency
about \( \omega =\omega _{i} \), encloses unit area and has a half
width of \( \sigma _{\omega }\sim 2\pi /T_{p} \). In the limit \(
T_{p}\to \infty  \) (while \( V_{p}^{2}T_{p} \) remains small)
this term can be taken as a \( \delta  \)-function expressing
precise energy conservation, so that only quasiparticles of energy
\( \hbar \omega  \) will be excited, i.e. \begin{eqnarray}
\label{EQN_ni_LONG_TIME} \lim _{T_{p}\to \infty }\langle \hat
n_{i}(T_p)\rangle &=&\frac{\pi V_{p}^{2}T_{p}N_{0}}{2}\left| \int
d\mathbf{r}(u^{*}_{i}+v^{*}_{i})e^{i\mathbf{q}\cdot
\mathbf{r}}|\psi
_{0}|\right| ^{2}\nonumber\\
&&\times\delta (\omega -\omega _{i})+\langle \hat n_{i}(0)\rangle
.
\end{eqnarray}

\section{Observable of Bragg
spectroscopy}\label{SECmeasurecond}

In this section we outline the experimental procedure of Bragg
spectroscopy
on condensates. We begin by discussing the measured observable, which
we refer to as the spectral response function. In \cite{Strenger99,
Stamper99}
this observable was assumed to be a measurement of the dynamic
structure factor. We briefly review the dynamic structure factor, and
discuss why it is inappropriate for these experiments.

\subsection{Spectral response function}

In the MIT experiments \cite{Stamper99, Strenger99} a low
intensity Bragg grating was used to excite the condensate
\emph{in-situ} for less than a quarter of a trap period.
Immediately following this the trap was turned off, the system
allowed to ballistically expand, and the momentum transfer to the
system was inferred by imaging the expanded spatial distribution.
The experimental signal measured is (see
\cite{Strenger99,Stamper99,Ketterle00})
\begin{equation}
\label{EQNfullspecrespfunc}
R(\mathbf{q},\omega )=\gamma \,\frac{P(T_p)-P(0)}{\hbar q},
\end{equation}
where\begin{equation}
\label{EQN_gamma_defn}
\gamma ^{-1}={\pi \over 2}N_{0}V_{p}^{2}T_{p},
\end{equation}
and $P$ is the momentum expectation of the system. We shall refer
to \( R(\mathbf{q},\omega ) \) as the \emph{spectral response function}.
For the case of a Gross-Pitaevskii wavefunction, the spectral response
function can be written as
\begin{equation}
\label{EQN_spec_resp_func}
R(\mathbf{q},\omega ) =\gamma \, \frac{\int d\mathbf{r}\, \Psi
^{*}(\mathbf{r},T_{p})\left( -i{\hbar \nabla}\right) \Psi
(\mathbf{r},T_{p})}{\hbar q},
\end{equation}
where it is assumed that the initial condensate has zero momentum
expectation value. The factors \( \gamma  \) and \( \hbar q \)
appearing in Eqs. (\ref{EQNfullspecrespfunc}) and
(\ref{EQN_spec_resp_func}) effectively scale out effects of the
Bragg intensity and duration, the magnitude of momentum transfer,
and condensate occupation, so that \( R(\mathbf{q},\omega ) \) can
be interpreted as a rate of excitation per atom (normalized with
respect to $V_p^2$) within the condensate.

\subsection{Dynamic structure factor}

The dynamic structure factor has played an important role in the
analysis of inelastic neutron scattering in superfluid $^{4}$He.
It has facilitated the understanding of collective modes, and has
enabled measurements of the pair distribution function and
condensate fraction in that system, as discussed extensively in
\cite{Griffin93}.
In those experiments a monochromatic neutron beam of momentum $\hbar \mathbf{%
k}_{0}$ is directed onto a sample of $^{4}$He and the intensity of neutrons
scattered to momentum $\hbar \mathbf{k}^{^{\prime }}$ is measured. van Hove %
\cite{LVH54} showed that the inelastic scattering cross section of thermal
neutrons, calculated in the first Born approximation, can be directly
expressed in terms of the quantity
\begin{eqnarray}\label{EQN_DSF_GENERAL}
S(\mathbf{q},\omega )&=&\frac{1}{\mathcal{Z}}\sum_{m,n}e^{-\beta
E_{m}}|\langle m|\hat{\rho}_{\mathbf{q}}|n\rangle |^{2}\delta (\hbar \omega
-E_{m}+E_{n}),  \BR
\end{eqnarray}%
which is called the dynamic structure factor (see \cite{Pines69}). In Eq. (%
\ref{EQN_DSF_GENERAL}), $|m\rangle $ and $E_{m}$ are the eigenstates and
energy levels of the unperturbed system, $\mathcal{Z}$ is the partition
function, $\hat{\rho}_{\mathbf{q}}=\int d\mathbf{r}\,\hat{\Psi}^{\dagger }(%
\mathbf{r})\exp (-i\mathbf{q}\cdot \mathbf{r})\hat{\Psi}(\mathbf{r})$ is the
density fluctuation operator, and we have taken%
\begin{eqnarray}
\hbar \omega  &=&\frac{\hbar ^{2}k_{0}^{2}}{2m}-\frac{\hbar ^{2}k^{\prime
2}}{2m},
\label{EQN_neutron_omega} \\
\mathbf{q} &=&\mathbf{k}_{0}-\mathbf{k}^{^{\prime }}.  \label{EQN_neutron_q}
\end{eqnarray}
Our choice of notation for these quantities is to facilitate comparison
between the matrix elements which arise in the dynamic structure factor and
Bragg cases. We note that in the Bragg context $\mathbf{q}$ and $\omega$ refer
to
the wave vector and frequency respectively of the optical potential, whereas
in a dynamic structure factor measurement, $\hbar \mathbf{q}$ and $\hbar
\omega$
are the momentum and energy respectively, transferred to the system from the
scattered probe.

For the case of a trapped gas Bose condensate, low intensity off
resonant inelastic light scattering \cite{Javanainen1995} provides
a close analogy to neutron scattering from $^{4} $He. Csord\'as
\emph{et al.} \cite{Csordas96} have shown that the cross section
for such inelastic light scattering, with energy and momentum
transfer to the photon of $-\hbar \omega$ and $-\hbar \mathbf{q}$
respectively, can be expressed in terms of the quantity
\begin{eqnarray}  \label{EQN_DSF_finite_temp}
S(\mathbf{q},\omega )&=&\sum _{i}\left| \int d\mathbf{r}\,
(u_{i}^{*}+v_{i}^{*})e^{i\mathbf{q}\cdot \mathbf{r}}|\psi _{0}|\right|
^{2}\nonumber\\
&&\times\left( (\langle \hat n_{i}\rangle +1)\delta (\omega -\omega _{i})+\langle
\hat n_{i}\rangle \delta (\omega +\omega _{i})\right).\BR
\end{eqnarray}
This expression generalizes the dynamic structure factor to the
case of light scattering and applies at finite temperatures in the
regime of linear response, where the Bogoliubov theory for
quasiparticles is valid. At $T=0$K, where thermal depletion can be
ignored, the dynamic structure factor (\ref{EQN_DSF_finite_temp})
takes the form
\begin{eqnarray}  \label{EQN_T0_DSF}
S_{0}(\mathbf{q},\omega )&=&\sum _{i}\left| \int d\mathbf{r}%
(u^{*}_{i}+v^{*}_{i})e^{i\mathbf{q}\cdot \mathbf{r}}|\psi _{0}|\right|
^{2}\delta (\omega -\omega _{i})\nonumber,\\
&&\qquad \qquad\qquad \qquad(T=0\rm{K})
\end{eqnarray}
for which a number of approximate forms have been calculated (e.g. see \cite%
{Csordas96,Wu96,Zambelli00}).

\subsubsection*{The dynamic structure factor and Bragg spectroscopy}
Although the dynamic structure factor and the spectral response
function are distinctly different quantities, they do resemble
each other strongly. In fact, the matrix elements in the $T=0$K
dynamic structure factor in (\ref{EQN_T0_DSF}) resemble those in
the expression (\ref{EQN_ni_LONG_TIME}) for the Bragg induced
quasiparticle population, $\langle \hat n_{i}\rangle $ in the long
time limit. It is easy to show, beginning from (\ref{EQN_T0_DSF}),
that
\begin{equation}  \label{EQN_DSF_in_terms_of_Total_Excit_Rate}
S_{0}(\mathbf{q},\omega )=\left\{ \gamma {\sum _{i}}\lim
_{T_{p}\to \infty }\langle \hat n_{i}(T_p)\rangle \right\} ,\qquad
(T=0\rm{K})
\end{equation}
where $\langle \hat n_{i}\rangle $ is given by Eq.
(\ref{EQN_ni_LONG_TIME}) and $%
\gamma $ is defined in Eq. (\ref{EQN_gamma_defn}). Note that because Eq. (%
\ref{EQN_DSF_in_terms_of_Total_Excit_Rate}) is evaluated at
$T=0$K, we have taken the initial occupation, $\langle \hat
n_{i}(0)\rangle $
in Eq.  (\ref%
{EQN_ni_LONG_TIME}), to be zero. In practice a very long pulse ($T_{p}\gg
1/\omega _{T} $) of sufficiently weak intensity ($V^{2}_{p}T_{p}\ll 1 $)
could be used to excite the quasiparticles in the regime necessary for Eq. (%
\ref{EQN_DSF_in_terms_of_Total_Excit_Rate}) to hold.
\begin{widetext}

In \cite{Stamper99, Strenger99} the spectral response function $
R(\mathbf{q},\omega ) $ was assumed to represent a measurement of the
dynamic structure factor, $S(\mathbf{q},\omega ).  $ The argument
given in those papers was based on the assumption that the Bragg pulse
would excite quasiparticles of definite momentum $\hbar \mathbf{q} $
and the momentum transfer is hence proportional to the rate of
quasiparticle excitation.  This is in fact not so, although we can
show that, in a certain sense, the dynamic structure factor and the
spectral response function do determine each other.
We use the result of Brunello \emph{et al.}
\cite{Brunello2001}, who  have shown that the momentum imparted can be related
to the dynamic structure factor according to
\begin{equation}  \label{EQN_Brunello_expression}
\frac{dP_{z}(t)}{dt}=-m\omega _{z}^{2}Z+2N_0\hbar q \left( \frac{V}{2}%
\right)^{2}\int d\omega '\, \left[ S(\mathbf{q},\omega
')-S(-\mathbf{q},-\omega ')\right] \frac{\sin ([\omega
-\omega ']t)}{\omega -\omega '},
\end{equation}
where the Bragg scattering has been taken to be in the $z $-direction and $%
P_{z} $ is the $z $-component of the momentum expectation.
The quantity $Z=\langle \sum^{N}_{j=1}z_{j}\rangle $ is the
expectation value of the $z $ center of mass coordinate and evolves
according to
\begin{equation}
\label{EQNZevolve}
\frac{dZ}{dt}=\frac{P_z}{m}.
\end{equation}
Taking the initial position and momentum expectations to be zero, Eqs.
(\ref{EQN_Brunello_expression}) and (\ref{EQNZevolve}) can be solved
using (\ref{EQNfullspecrespfunc}) to give the spectral response
function in terms of the dynamic structure factor as
\begin{equation}
\label{EQNsrffrommtmexptn}
R(\mathbf{q},\omega)=\frac{1}{\pi T_p}
\int d\omega^\prime\left[S(\mathbf{q},\omega^\prime)-
S(-\mathbf{q},-\omega^\prime)\right]
\frac{\cos(\omega_zT_p)-\cos([\omega-\omega^\prime]T_p)}{[\omega-\omega^
\prime]^2-\omega_z^2}.
\end{equation}
This formula can be inverted, but the inversion formulae are different
depending on whether $\omega_z $ is zero or not:
\begin{eqnarray}\label{EQNnotrapSRFlongtime}
S(\mathbf{q},\omega)-S(-\mathbf{q},-\omega)
&=& \lim_{T_p\to\infty} R(\mathbf{q},\omega,T_p)
                          \qquad  \qquad\quad\mbox{for $ \omega_z=0$},
\\\label{EQNnotrapSRFlongtime1}
&=& \omega_z^2\int_0^\infty R({\bf q},\omega,T_p)\,T_pdT_p
                            \qquad  \qquad\quad\mbox{for $ \omega_z\ne0$}.
\end{eqnarray}
In the linearized approximation we are using, we can use
(\ref{EQN_DSF_finite_temp}) to show that
\begin{eqnarray}\label{zero}
S_0(\mathbf{q},\omega)&=&S(\mathbf{q},\omega)-S(-\mathbf{q},-\omega),\qquad
\omega\ge0,
\end{eqnarray}
so that the differencing cancels out finite temperature effects.  Beyond the
Bogoliubov approximation, there will however be residual finite-temperature
effects.
\end{widetext}

Thus we see it is possible to determine the {\em zero temperature}
dynamic structure factor in a good degree of approximation from
experiments on a trapped condensate, provided measurements are
performed for a sufficient range of pulse times $ T_p$.  It is
clear that this could be a difficult experiment to implement,
since the spectral response function will drop off faster than $
1/T_p$ for large $ T_p$, so the measured signal could become very
small. But in analogy with the results of
Sect.\ref{SEC_long_time}, we would expect that significant
information could be obtained by measuring for pulse times $ T_p$
up to about 5 trap periods.

On the other hand there is no direct connection between the
spectral response function and the dynamic structure factor for
any {\em single} time $ T_p$ unless $ \omega_z=0$, in which case
we must take $ T_p\to \infty$. In fact, if we note that
significant structure in $ S({\bf q},\omega)$ is expected on the
frequency scale of $ \omega_z$, then the formula
(\ref{EQNsrffrommtmexptn}) shows that a smearing over this
frequency scale is assured by the form of the integrand in
(\ref{EQNsrffrommtmexptn}), independently of the value of $ T_p$.

One therefore must conclude that the spectral response function, not
the dynamic structure factor, is the appropriate method of analysis
for Bragg scattering experiments.  However, it is in principle
possible to determine the {\em zero temperature} dynamic structure
factor in a certain degree of approximation from these experiments by
use of the inversion formula (\ref{EQNnotrapSRFlongtime1}).

\section{Quasi-homogeneous approximation to \protect\(
R(\mathbf{\lowercase{ q}},\omega )\protect \)}\label{SEC_QH}

In this section we develop an approximation for the spectral response
function valid for time scales shorter than a quarter trap period.
Because trap effects are negligible on this time scale we employ a
quasi-homogeneous approach, based on homogeneous quasiparticles
weighted
by the condensate density distribution. This approximation plays an
important part in the analysis of the numerical results we present
in section \ref{SECBraggspec}.

\subsection{Homogeneous spectral response}

The results developed in section \ref{SEC_THEORY} for the
trapped condensates can be applied to a homogeneous system of number density
$n$
by making the replacements \begin{eqnarray}
u_{i}(\mathbf{r}) & \rightarrow  & u_{k}(n)e^{i\mathbf{k}\cdot
\mathbf{r}}/\sqrt{\mathcal{V}},\label{EQN_uQH} \\
v^{*}_{i}(\mathbf{r}) & \rightarrow  & v_{k}(n)e^{-i\mathbf{k}\cdot
\mathbf{r}}/\sqrt{\mathcal{V}},\label{EQN_vQH} \\
\sqrt{N_{0}}\psi _{0}(\mathbf{r}) & \rightarrow  &
\sqrt{N_{0}/\mathcal{V}}=\sqrt{n,}\label{EQN_psiQH}
\end{eqnarray}
 where \( \mathcal{V} \) is the volume, and \begin{eqnarray}
u_{k}(n) & = & \frac{\omega ^{B}_{k}(n)+\omega _{k}}{2\sqrt{\omega
^{B}_{k}(n)\omega _{k}}}\, ,\label{EQN_Bog_u} \\
v_{k}(n) & = & -\frac{\omega ^{B}_{k}(n)-\omega _{k}}{2\sqrt{\omega
^{B}_{k}(n)\omega _{k}}}\, ,\label{EQN_Bog_v} \\
\omega ^{B}_{k}(n) & = & \sqrt{\omega ^{2}_{k}+2nU_{0}\omega
_{k}/\hbar }\, ,\label{EQN_BOG_disp}\\
\omega_{k} & = & \frac{\hbar k^2}{2m}\, .
\end{eqnarray}
We have explicitly written these as functions of density $n$ for
later convenience.
 Using Eqs.
(\ref{EQN_uQH})-(\ref{EQN_psiQH}) the expectation of the
quasiparticle operators (\ref{EQN_bi_vac_expt_finite_Tp}) (i.e. at
\( T=0 \)K) resulting from Bragg excitation is found to be
\begin{widetext}
\begin{eqnarray}
\langle \hat{b}_{\mathbf{k}}\rangle  & = &
-iV_{p}\sqrt{\frac{n}{\mathcal{V}}}\int
_{\mathcal{V}}d\mathbf{r}\, (u_{k}(n)e^{-i\mathbf{k}\cdot
\mathbf{r}}+v_{k}(n)e^{-i\mathbf{k}\cdot \mathbf{r}})e^{i\omega
^{B}_{k}(n)T_{p}/2}\label{EQN_bk_finite_time}
 \Big [e^{-i\omega T_{p}/2}\left( \frac{\sin ((\omega
^{B}_{k}(n)-\omega )T_{p}/2)}{(\omega ^{B}_{k}(n)-\omega )}\right)
e^{i\mathbf{q}\cdot \mathbf{r}}\nonumber \\
 &  & \quad +\, e^{i\omega T_{p}/2}\left( \frac{\sin ((\omega
^{B}_{k}+\omega )T_{p}/2)}{(\omega ^{B}_{k}+\omega )}\right)
e^{-i\mathbf{q}\cdot \mathbf{r}}\Big ].
\end{eqnarray}
Since the homogeneous excitations are plane waves, evaluating the
spatial integral in Eq. (\ref{EQN_bk_finite_time}) selects
out quasiparticles with wave vector \(
\mathbf{k}=\pm \mathbf{q} \).
The occupations \textbf{\( \langle \hat{b}_{\mathbf{k}}^{\dagger
}\hat{b}_{\mathbf{k}}\rangle  \)}
of these states are\begin{equation}
\label{EQN_Bog_Pop}
\langle \hat{n}_{\pm \mathbf{q}}\rangle =\frac{\pi
V_{p}^{2}T_{p}N_{0}}{2}(u_{q}(n)+v_{q}(n))^{2}F((\omega ^{B}_{q}(n)\mp \omega
),T_{p}),
\end{equation}
 where \( F(\omega ,T) \) is defined in Eq.
(\ref{EQN_time_dep_pert_F}).
\end{widetext}
Since a quasiparticle created by \( \hat{b}_{\mathbf{k}}^\dagger \) carries
momentum \( \hbar \mathbf{k} \), the total momentum transferred to
the homogeneous condensate is \begin{equation}
\label{EQN_homog_mtm_trans}
\langle \mathbf{p}\rangle\ =\hbar \mathbf{q}\left( \langle
\hat{n}_{\mathbf{q}}\rangle -\langle \hat{n}_{-\mathbf{q}}\rangle
\right) ,
\end{equation}
and the spectral response function, as defined in Eq.
(\ref{EQNfullspecrespfunc}), becomes
\begin{eqnarray}
&&
R(\mathbf{q},\omega )  =  \gamma \,  \frac{\left|\langle \mathbf{p}\rangle
\right|
}{\hbar q},
\\
 &&\, =  \gamma \, \left( \langle \hat{n}_{\mathbf{q}}\rangle -\langle
\hat{n}_{-\mathbf{q}}\rangle \right) ,\label{EQN_srf_QN1}
\\
 &&\, =  \frac{\omega _{q}}{\omega ^{B}_{q}(n)}\Big[ F((\omega
^{B}_{q}(n)-\omega ),T_{p})
-F((\omega ^{B}_{q}(n)+\omega ),T_{p})\Big]
,\BR\label{EQN_HSRF}
\end{eqnarray}
where we have used \( (u_{q}(n)+v_{q}(n))^{2}=\omega _{q}/\omega ^{B}_{q}(n)
\)
(from Eqs. (\ref{EQN_Bog_u}) and (\ref{EQN_Bog_v}))
to arrive at the last result.

\subsection{Quasi-homogeneous spectral response function}

The density distribution of a Thomas-Fermi condensate is given
by\begin{equation}
\label{EQN_TF_density_dist}
N(n)=\frac{15N_{0}}{4n^{2}_{p}}n\sqrt{1-n/n_{p}},
\end{equation}
 where \( N(n)dn \) is the portion of condensate atoms in the density
range \( n\rightarrow n+dn \), and \( n_{p}=\hbar \mu /U_{0} \),
is the peak density (see \cite{Strenger99}).

To approximate the spectral response function for the inhomogeneous
case we multiply the portion of the condensate at density \( n \)
by the homogeneous spectral response function (\ref{EQN_HSRF})
for a homogeneous condensate (of density \( n \)) and integrate over
all densities present, i.e. \begin{eqnarray}
R_{QH}(\mathbf{q},\omega ) & \equiv & \int dn\, N(n)R(\mathbf{q},\omega
),\label{EQN_srf_QH2} \\
 & = & \left( \frac{15N_{0}}{4n^{2}_{p}}\right) \int ^{n_{p}}_{0}dn\,
\frac{n\,\omega _{q}}{\omega
^{B}_{q}(n)}  \sqrt{1-n/n_{p}}\label{EQN_quasihomo_finitetime_srf}
\nonumber\\
 &  & \times \Big [F(\omega ^{B}_{q}(n)-\omega ,T_{p})-F(\omega
^{B}_{q}(n)+\omega,T_{p})\Big ].\BR
\end{eqnarray}
 We shall refer to \( R_{QH} \) as the \emph{finite time
quasi-homogeneous
approximation} to the spectral response function, or simply the
\emph{quasi-homogeneous
approximation.} Ignoring the finite time broadening effects and
assuming
exact energy conservation in \( R_{QH} \) with the replacement \(
F(\omega, T)\rightarrow \delta (\omega ) \),
reduces Eq. (\ref{EQN_quasihomo_finitetime_srf}) to the simpler
line-shape expression
\begin{eqnarray}
R_{QH}(\mathbf{q},\omega )&\rightarrow& I_{q}(\omega ),
\\
&=&\frac{15\hbar (\omega ^{2}-\omega ^{2}_{q})}{8\omega
_{q}N_{0}U_{0}}\sqrt{1\!-\!\frac{\hbar (\omega ^{2}-\omega
^{2}_{q})}{2\omega _{q}N_{0}U_{0}}} \label{EQN_LDA},
\end{eqnarray}
where we have adopted the notation, \( I_{q} \), as used in the
original derivation \cite{Strenger99}. Equation (\ref{EQN_LDA}) is
also known as the local density approximation to the dynamic
structure factor (see \cite{Zambelli00}), and has been used to
analyze experimental data in \cite{Strenger99,Stamper99}. We
emphasize that with $T_p < (1/4)\,T_{ \rm{trap}}$ the
$\delta$-function replacement is unjustifiable, as we verify with
our numerical results in the next section.

\section{Bragg spectroscopy}\label{SECBraggspec}

Bragg spectroscopy can broadly be defined as selective excitation of
momentum components in a condensate, by Bragg light fields. In this
section we consider
the spectral response function as a Bragg spectroscopic measurement,
and using numerical simulations of the Gross-Pitaevskii equation
(\ref{EQN_TD_GPE}) and
the analytic results of the previous section we identify the
dominant physical mechanisms governing the spectral response
behavior.
We investigate the spectral response function for a vortex and identify
parameter regimes in which a clear signature of a vortex
is apparent. From the full numerical simulations we are also able
to assess the effect
of higher laser intensities on the spectroscopic measurements.

\subsection{Numerical results for \protect\( R(\mathbf{q},\omega
)\protect \)\label{SEC_spec_resp_func} }

\subsubsection{Procedure}

The numerical results we present for \( R(\mathbf{q},\omega ) \),
are found by evolving an initial stationary condensate state
(typically a ground state) in the presence of the Bragg optical
potential, using Eq. (\ref{EQN_TD_GPE}). At the conclusion of this
pulse, the spectral response is evaluated using Eq.
(\ref{EQN_spec_resp_func}). This differs slightly from the typical
procedure in the experiments, where the system is allowed to
expand before destructive imaging is used to measure the
condensate momentum. However we have verified numerically (in
cylindrically symmetric 3D cases) that condensate expansion (after
the pulse) does not alter the momentum expectation value. For each
desired value of \( \mathbf{q} \) and \( \omega  \), we repeat our
procedure of evolving \( \Psi  \) according to (\ref{EQN_TD_GPE}),
and calculating \( R(\mathbf{q},\omega ) \) immediately after the
optical pulse terminates.

For axially symmetric situations, the simulations are calculated in
three spatial dimensions with \( \mathbf{q} \) oriented along the
\( z \)-axis. When the initial state is a vortex, the interesting
case of scattering in a direction orthogonal to the vortex core
(lying on the $z$-axis) would
break the symmetry requirement, so for these cases 2D simulations
with \( \mathbf{q} \) directed along the \( y \)-axis are used.
For convenience we use computational units of distance \(
r_{0}=\sqrt{\hbar /2m\omega _{T}} \);
interaction strength \( w_{0}=\hbar \omega _{T}r_{0}^{3} \); and
time \( t_{0}=\omega _{T}^{-1} \) ; where \( \omega _{T} \) is the
trapping frequency in the direction of scattering.

\subsubsection{Parameter regimes}

We use square pulses of intensity and duration such that typically
less than \( 1\% \) of the condensate is excited, except in section
\ref{nlresp} where we investigate the nonlinear response of the condensate.
As long as the amount of excitation is small, we verify that the spectral
response function $R({\bf q},\omega)$ is independent of \( V_{p} \).
However, the shape of $R({\bf q},\omega)$ is dependent on the pulse
duration (in accordance with the frequency spread about \( \omega  \)
associated with the time limited pulse) and on the magnitude of
\( \mathbf{q} \).

The momentum \( \hbar q_{0} \) defined by \begin{equation}
\label{EQN_q0}
\hbar q_{0}=\sqrt{2mn_{p}U_{0}},
\end{equation}
(i.e. \( q_{0}=1/\xi  \), where \( \xi  \) is the condensate healing
length) characterizes the division between regimes of phonon and free
particle-like quasiparticle character. We note that the experimental
results in \cite{Strenger99} and \cite{Stamper99} report measurements
of \( R(\mathbf{q},\omega ) \) in the free particle and phonon regime,
respectively.

\subsection{Underlying broadening mechanisms}

\begin{figure}
{\centering \includegraphics{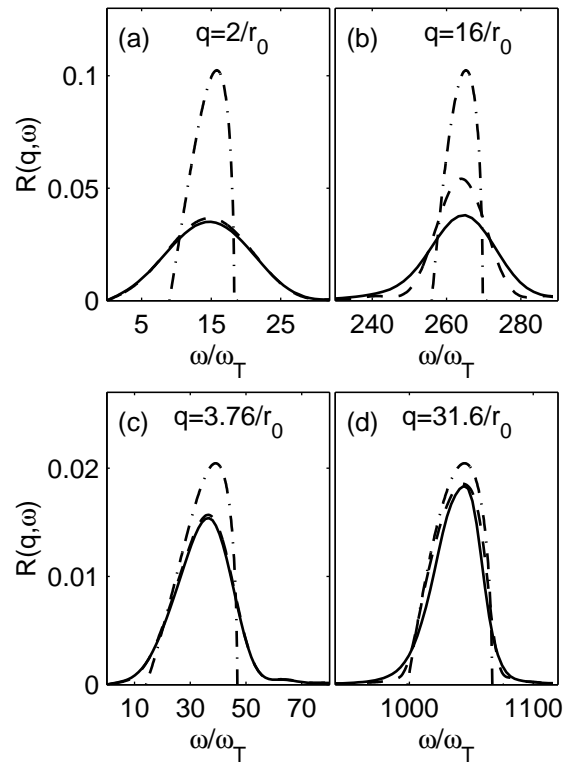}
\par}

\caption{\label{FIG_spec_resp_comp} Spectral response function
\protect\( R(\mathbf{q},\omega )\protect \) of a 3D condensate.
(a), (b) Spherical condensate, \protect\(
N_{0}U_{0}=10^{4}w_{0}\protect \), \protect\( \mu =14.2\omega
_{T}\protect \). (c), (d) Oblate condensate with trap asymmetry
\protect\( \lambda =\sqrt{8}\protect \), \protect\(
N_{0}U_{0}=5.6\times 10^{5}w_{0}\protect \), \protect\( \mu
=70.9\omega _{T}\protect \). Full numerical solution for
\protect\( R\protect \) shown as solid line; the local density
approximation (\protect\( I_{q}\protect \)) dash-dot; and the
finite time local density approximation (\protect\( R_{QH}\protect
\)) dashed. Bragg parameters are \protect\( V_{p}=0.2\omega
_{T}\protect \) and \protect\( T_{p}=0.4/\omega _{T}.\protect \) }
\end{figure}

We present in Fig. \ref{FIG_spec_resp_comp}  spectral response
functions calculated using the Gross-Pitaevskii simulations in
three spatial dimensions. The spectral response functions in Figs.
\ref{FIG_spec_resp_comp}(a) and (b) are for a spherically
symmetric ground state with \( q_{0}=3.8/r_{0}. \) The \( q \)
values (of the Bragg fields) were chosen so that the case in Fig.
\ref{FIG_spec_resp_comp}(a) is in the phonon regime (\( q=2/r_{0}
\)), while Fig. \ref{FIG_spec_resp_comp}(b) is in the free
particle limit (\( q=16/r_{0} \)). For comparison, we present in
Figs. \ref{FIG_spec_resp_comp}(c) and (d) spectral response
functions for a different ground state of greater nonlinearity,
for which \( q_{0}=8.4/r_{0} \). Fig. \ref{FIG_spec_resp_comp}(c)
is in the phonon regime (\( q=3.76/r_{0} \)), whereas Fig.
\ref{FIG_spec_resp_comp}(d) is in the free particle regime (\(
q=31.6/r_{0} \)). In all cases we compare the Gross-Pitaevskii
calculation of \( R \) with the lines-shape \( I_{q} \) and
quasi-homogeneous result \( R_{QH} \) from Eqs. (\ref{EQN_LDA})
and (\ref{EQN_quasihomo_finitetime_srf}) respectively.

\subsubsection{Mechanisms}
The two ground states used in calculating the results presented in
Fig.  \ref{FIG_spec_resp_comp} are both in the Thomas-Fermi limit
(i.e.  they satisfy the condition \( \mu \gg \omega _{T}
\))\footnote[2]{For our numerical simulations we use
Gross-Pitaevskii eigenstates calculated from Eq.
(\ref{EQN_TIGPE})}.  We can see from the figure that the local
density approximation \( I_{q} \) Eq. (\ref{EQN_LDA}) used by
previous authors does not always give a good description of \(
R(\mathbf{q},\omega ) \), whereas our quasi-homogeneous
approximation \( R_{QH}(\mathbf{q},\omega ) \) Eq.
(\ref{EQN_quasihomo_finitetime_srf}) is much more accurate.  We
have investigated the accuracy of \( R_{QH} \) over a wide
parameter range which has allowed us to evaluate the relative
importance of the three underlying broadening mechanisms which
contribute to \( R(\mathbf{q},\omega ) \). The mechanisms and
their contributions are as follows:
\begin{itemize}
\item[i)] The shift in the excitation
spectrum due to the meanfield interaction (depending on the local
density). The range of densities present in the condensate cause
a spread in this shift. The frequency width
associated with this spread is proportional to the
chemical potential \( \mu  \), and we shall refer to this as the density
width  - see section \ref{SEC_QH}.

\item[ii)] The Doppler effect due to the momentum spread of the condensate;
the Doppler broadened frequency width is \( \sigma _{p}\cdot q/m
\), where \( \sigma _{p} \) is the condensate momentum width.
\item[iii)]
The frequency spread in the Bragg grating due to the finite pulse
time; the width arising from this effect is \( \sim \pi/T_{p} \).
\end{itemize}
\subsubsection{Relative importance of mechanisms}
The relative importance of these mechanisms varies according to the
parameter regime. In Table \ref{TAB_groundstatesmechanisms}
we compare the estimated values of these widths for the simulations
in Fig. \ref{FIG_spec_resp_comp}.
\begin{table}[!tbh]
\begin{center}

\begin{tabular}{|l|c|c|c|}
\hline
&
density&
finite \( T_{p} \)&
momentum \\
\hline Fig.& \( \mu   \) & \( \pi /T_{p}  \)&
\( \sigma _{p}\cdot q/m  \)\\
\hline
\hline
\ref{FIG_spec_resp_comp}(a)&
14.2&
7.9&
0.86\\
\hline
\ref{FIG_spec_resp_comp}(b)&
14.2&
7.9&
6.92\\
\hline
\ref{FIG_spec_resp_comp}(c)&
70.6&
7.9&
0.88\\
\hline
\ref{FIG_spec_resp_comp}(d)&
70.6&
7.9&
7.4\\
\hline
\end{tabular}

\end{center}

\caption{\label{TAB_groundstatesmechanisms} Simple estimate of
widths of component mechanisms of the spectral response functions
in Fig. \ref{FIG_spec_resp_comp}. All quantities expressed in
units of $\omega_T$.}
\end{table}
 We see that for the case of Fig. \ref{FIG_spec_resp_comp}(a)
both finite time and density effects are important; hence the
quasi-homogeneous
result which includes both of these is in good agreement with \(
R(\mathbf{q},\omega ) \),
whereas \( I_{q}(\omega ) \) which accounts for density effect alone
is quite inaccurate. In Fig. \ref{FIG_spec_resp_comp}(b) the
value of \( q \) is much larger, increasing the momentum
width to a point where it is comparable to the other mechanisms. Since
both \( I_{q} \) and \( R_{QH} \) fail to account for the condensate
momentum, neither approximation to the spectral response function
is in particularly good agreement with the numerically calculated
\( R \).
In both Figs. \ref{FIG_spec_resp_comp}(c)
and (d) the density width is an order of magnitude larger than both
the finite pulse and momentum widths, and so in this regime the
simple line-shape expression \( I_{q} \) is generally adequate.

\subsubsection{Experimental comparison}

Taking typical experimental parameters, the state used in Figs.
\ref{FIG_spec_resp_comp}(a) and (b) corresponds to about \(
2.8\times 10^{5} \) Na atoms in a \( 50 \) Hz trap, with the Bragg
grating formed from \( 589 \) nm laser beams intersecting at an
angle of \( 5^{\circ } \) in (a) and \( 39^{\circ } \) in (b).
These figures are indicative of typical Bragg spectroscopy results
for a small to medium size condensate.  The state used in Figs.
\ref{FIG_spec_resp_comp}(c) and (d) corresponds to about \( 10^{7}
\) Na atoms in a \( 100 \) Hz trap, and was chosen to match some
of the features of the experiments reported in \cite{Strenger99,
Stamper99}, e.g.  the peak density of this state is \( \sim
4\times 10^{14} \) atoms/cm\(^{3} \) and the chemical potential is
\( \sim 6.7 \)kHz.  The momentum values chosen correspond to those
used to probe the phonon and free particle regimes in those papers
(Bragg grating formed by \( 589 \) nm beams at \( 14^{\circ } \)
and \( 180^{\circ } \) respectively).  Computational constraints
mean we cannot match the experimental (prolate) trap geometry, but
in a similar manner to the measurements made in \cite{Strenger99}
we scatter along a tightly trapped direction, although from a
condensate in an oblate trap of aspect ratio \( \sqrt{8} \). We
note that in the phonon probing experiments \cite{Stamper99},
scattering was performed in the weakly trapped direction for
imaging convenience.  It is worth emphasizing that the mechanisms
accounted for in our approximate response function \( R_{QH} \)
depend only on the peak density (i.e.  \( \hbar \mu /U_{0} \)),
the magnitude of \( q \) and the Bragg pulse duration. The reason
is that condensates in the Thomas-Fermi regime with the same peak
density will have identical density distributions in either
prolate or oblate traps (see Eq. (\ref{EQN_TF_density_dist})).
Thus the quasi-homogeneous approximation predicts the same
spectral response function for both.  Scattering in a tightly
trapped direction will enhance momentum effects not accounted for
in \( R_{QH} \), since spatially squeezing the condensate causes
the corresponding momentum distribution to broaden.  However, it
is apparent from Table \ref{TAB_groundstatesmechanisms}, that for
the case of large condensates this momentum effect is relatively
small, and thus we expect our result in Figs.
\ref{FIG_spec_resp_comp}(c) and (d) to give a reasonably accurate
description of the MIT experiments \cite{Strenger99, Stamper99}.

\subsection{Spectral response function of a vortex}

\begin{figure}

{\centering \includegraphics{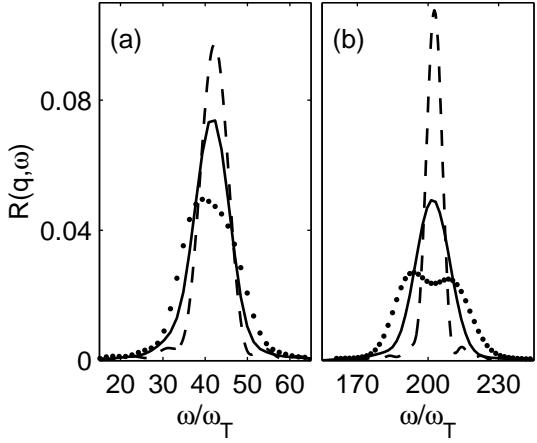}
\par}

\caption{\label{FIG_vrtx_grnd_srf_profiles} Spectral response
functions
\protect\( R(\mathbf{q},\omega )\protect \) for 2D ground (solid)
and \protect\( m=1\protect \) vortex (dots) states, and \protect\(
R_{QH(2D)}\protect \)
for 2D ground state (dashed). Both states have \protect\(
N_{0}U_{0}=500w_{0}\protect \),
and the spectra are calculated with (a) \protect\( q=6/r_{0}\protect
\)
and (b) \protect\( q=14/r_{0}\protect \). Bragg parameters are
\protect\( V_{p}=0.2\omega _{T}\protect \)
and \protect\( T_{p}=0.8/\omega _{T}\protect \).}
\end{figure}

In  previous work \cite{Blakie01}
we showed how Bragg scattering from a vortex can produce
an asymmetric spatially selective beam of scattered
atoms which provides an \emph{in-situ} signature of a vortex. In Fig.
\ref{FIG_vrtx_grnd_srf_profiles} we compare the spectral
response
functions from two dimensional ground and vortex states, and in Table
\ref{TAB_vort_mechanisms} we summarize the density, finite
time and momentum effects for the cases in Fig.
\ref{FIG_vrtx_grnd_srf_profiles}.
\begin{table}[!tbh]
\begin{center}

\begin{tabular}{|l|c|c|c|}
\hline
&
density&
finite time&
momentum\\
\hline \hline \multicolumn{1}{|c|}{Fig.}& \( \mu  \) & \( \pi
/T_{p} \)&
\( \sigma _{p}\cdot q/m  \)\\
\hline
\ref{FIG_vrtx_grnd_srf_profiles}(a) ground&
9.0&
3.9&
2.9\\
\hline
\ref{FIG_vrtx_grnd_srf_profiles}(a) vortex&
9.2&
3.9&
5.2\\
\hline
\ref{FIG_vrtx_grnd_srf_profiles}(b) ground&
9.0&
3.9&
6.7\\
\hline
\ref{FIG_vrtx_grnd_srf_profiles}(b) vortex&
9.2&
3.9&
12.2\\
\hline
\end{tabular}

\end{center}

\caption{\label{TAB_vort_mechanisms} Simple estimate of widths of
component mechanisms of the spectral response functions in Fig.
\ref{FIG_vrtx_grnd_srf_profiles}. All quantities expressed in
units of $\omega_T$.}
\end{table}
 In the low momentum transfer case Fig.
\ref{FIG_vrtx_grnd_srf_profiles}(a),
the density width is the most significant component of the response
function width, although the increased momentum distribution of the
vortex state relative to the ground state is reflected in the width
of \( R \). We also see that \( R_{QH} \) for the 2D ground state
is a reasonable approximation to the full numerical calculation of
\( R \).

In the large momentum transfer case Fig.
\ref{FIG_vrtx_grnd_srf_profiles}(b),
Doppler effects (which scale linearly with \( q \)) become important
and the two peaks in the vortex spectral response appear. These
indicate
the presence of flow (or momentum components) running parallel and
anti-parallel to the scattering direction.
Zambelli \emph{et al.}  \cite{Zambelli00} have discussed the Doppler effect
in detail, and they utilized the \emph{impulse approximation} for the dynamic
structure factor for the regime where this mechanism dominates. The
essence of this approximation is to project the trapped condensate
momentum distribution into frequency space, and they have applied this
to the case of a vortex (in the non-interacting long pulse limit). The impulse
approximation always predicts a double peaked response from a vortex state
corresponding to the Doppler resonant frequencies for the parallel
and anti-parallel momentum components. Our numerical results for
$R(\mathbf{q},\omega)$ show that at low momenta this vortex signature is
obscured
by the density and finite time effects.

\subsection{Shape characteristics of the spectral response
function}

The spectral response functions shown in Figs.
\ref{FIG_spec_resp_comp}
and \ref{FIG_vrtx_grnd_srf_profiles} contain a large amount
of information, however the major properties can be well represented
by a few numbers that describe the overall shape characteristics of
these curves. The characteristics of the response curve that we focus
on are the area under the curve, the shift of the mean response
function
and the rms frequency width. These can be expressed in terms of the
first few frequency moments (\( m_{n} \)) of \( R \) at constant
\( \mathbf{q} \), defined as \begin{equation}
\label{EQN_n_moment}
m_{n}\equiv\int d\omega \, \omega ^{n}R(\mathbf{q},\omega ).
\end{equation}
The area under the curve is simply \( m_{0} \); the shift of mean
response frequency is \begin{equation}
\label{EQN_freq_shift}
m_{\rm {shift}}=m_{1}/m_{0}-\omega _{q};
\end{equation}
 and the rms response width is \begin{equation}
\label{EQN_freq_rms_width}
m_{\rm {rms}}=\sqrt{m_{2}/m_{0}-(m_{1}/m_{0})^{2}}.
\end{equation}
These characteristics have also been considered in experimental
measurements \cite{Stamper99, Strenger99}, where, however, only two
values of \( q \) were used (corresponding to Bragg gratings formed by
counter-propagating and approximately co-propagating beams
respectively) and the speed of sound (equivalent to \( q_{0} \)) was
changed by altering the condensate density.

Numerically we can consider any value \( q \) which can be resolved on
our computational grids.  Here we choose for simplicity to directly
vary \( q \) and consider how the spectral response function changes.
The initial state for these calculations is identical to that used in
Figs.  \ref{FIG_spec_resp_comp}(a)-(b) (i.e.  a spherically symmetric
ground state with \( N_{0}U_{0}=10^{4}w_{0} \)), because density,
temporal and momentum effects are all important for the range of \( q
\) values we consider.  This should be contrasted with the initial
state in Figs.  \ref{FIG_spec_resp_comp}(c)-(d), where the density
effect dominates over the other two effects at both small and large \(
q \).

\begin{figure}

{\centering \includegraphics{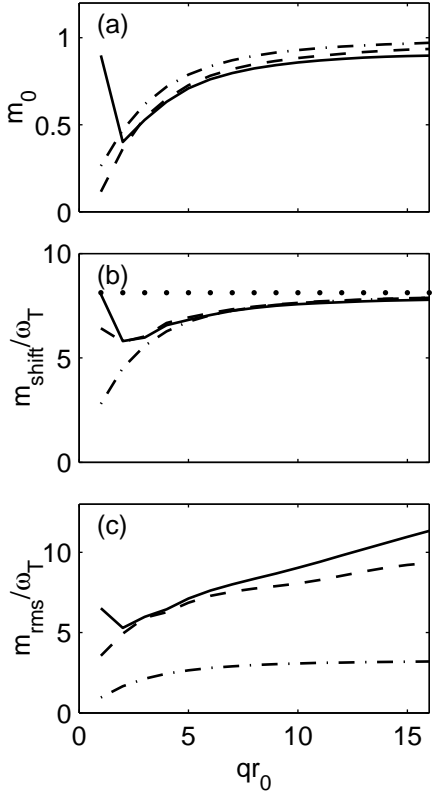} \par}

\caption{\label{FIG_srf_moments} Spectral response function
properties
(see text) for a 3D condensate with \protect\(
N_{0}U_{0}=10^{4}w_{0}\protect \).
\protect\( R(\mathbf{q},\omega )\protect \) (solid), \protect\(
R_{QH}(\mathbf{q},\omega )\protect \)
(dashed) and \protect\( I_{q}(\omega )\protect \) (dash-dot). (a)
the zeroth moment; (b) the mean response shift, with the shift
\protect\( 4\mu /7\protect \)
indicated (dotted); (c) the rms width of the frequency response. Bragg
parameters are \protect\( V_{p}=0.2\omega _{T}\protect \) and
\protect\( T_{p}=0.4/\omega _{T}\protect \).}
\end{figure}

Our numerical results lead to the following observations:
\begin{itemize}
\item[i)]
In Fig. \ref{FIG_srf_moments}(a) the area under the response curve
(i.e. \( m_{0} \)) is considered. The reduction of this from unity
is characteristic of suppression of scattering for \( q<q_{0} \)
caused by the interference of the \( u \) and \( v \) amplitudes
in the expression for the quasiparticle population (e.g. see Eq.
(\ref{EQN_Bog_Pop})). Apart from the discrepancy at very low \( q
\) (which exists in all moments and which we discuss more fully
below), both \( I_{q} \) and \( R_{QH} \) are in qualitatively
good agreement with \( R \). The finite pulse duration effect on
\( m_{0} \) is to reduce it uniformly from the \( I_{q} \)
prediction, a feature which \( R_{QH} \) represents well.

\item[ii)]
The mean shift (\ref{EQN_freq_shift}) arises because of the
Hartree interaction between the particles excited and those remaining
in the condensate, as indicated by the Bogoliubov dispersion relation
(\ref{EQN_BOG_disp}). The approximations \( R_{QH} \) and
\( I_{q} \) are seen to be in good agreement with the full numerical
calculation over most of the range of \( q \), apart from \( q=0 \)
(see below). At large values of \( q \), the mean shift saturates
at the value \( 4\mu /7 \).

\item[iii)]
The width prediction of \( R_{QH} \) is in good agreement with the
full numerical calculation at low \( q \) (\( 2\le qr_{0}\lesssim
6 \)), but differs as \( q \) increases and the role of the
condensate momentum distribution increases in importance. \( I_{q}
\) is in poor agreement for all \( q \), because it ignores
temporal broadening effects.

\item[iv)] In Fig. \ref{FIG_srf_moments} the sharp
features in the full numerical calculation of the moments of \( R \)
at extremely low \( q \) arise because the (ground state) condensate
momentum wavefunction has significant density in the region where
the majority of the condensate is being scattered. This causes
stimulated
scattering of the condensate atoms, and will significantly affect
the spectral response function when the amount of condensate initially
present in the region we are scattering to is similar (or larger)
than the amount of condensate excited with the Bragg pulse. This effect
has been ignored in the derivation of $R_{QH}$ by taking $\hat b_\mathbf{k}$
to have a zero initial occupation in Eq. (\ref{EQN_bk_finite_time}).
\end{itemize}

\subsubsection*{Evaluating condensate momentum width from the spectral
response function }

The MIT group have used the rms-width of the spectral response function to
determine the Doppler width $\Delta \omega _{\mathrm{Dop}}$ of the
condensate, and hence the momentum width $\sigma _{p}(=(m/q)$ $\Delta \omega
_{\mathrm{Dop}}).$ This allowed them to make the important observation that
the  condensate coherence length is at least as large as the condensate size %
\cite{Strenger99} (also see \cite{Ketterle00}). The Doppler width was
obtained from the spectral response profile by assuming that the Doppler
effect adds in quadrature to the density and temporal effects. The precise
details of the procedure used in \cite{Strenger99} are not given (see Fig. 3
of \cite{Strenger99}), however an expression for the momentum width of the
form
\begin{equation}
\sigma _{p}^{^{\prime }}=\frac{m\,\sqrt{m_{\mathrm{rms}}^{2}-(\Delta
I_{q})^{2}-(\Delta F(\omega ,T_{p}))^{2}}}{q},  \label{EQN_MIT_quad}
\end{equation}%
is implied in the text.  A difficulty that arises in applying Eq. (\ref%
{EQN_MIT_quad})  is that the rms-frequency spread $\Delta F(\omega ,T_{p})$
due to the finite pulse length is ill defined. Formally $F(\omega ,T_{p})$
has an unbounded rms-width, though the central peak has a half width of $\pi
/T_{p}$. In order to allow the best possible result from the MIT procedure,
we have treated $\Delta F$ as an arbitrary parameter, and fitted the
function  $\sigma _{p}^{^{\prime }}$ in Eq.(\ref{EQN_MIT_quad}) to the
actual condensate momentum width of the  ground state we use. The result for
a range of $q$ values is shown plotted as circles in Fig. \ref%
{FIG_momentum_width}, and was obtained with a best fit value
$\Delta F\approx 0.86\times\pi /T_{p}$. We can see that Eq.
(\ref{EQN_MIT_quad}) does not give a particularly good estimate of
the true momentum width, indicating that the assumption of
quadrature contributions of density, momentum and finite time
effects to the spectral response function  is inappropriate.
Agreement improves as $q$ increases, as would be expected, because
the Doppler effect dominates over the other mechanisms in that
regime.

\begin{figure}
{\centering \includegraphics{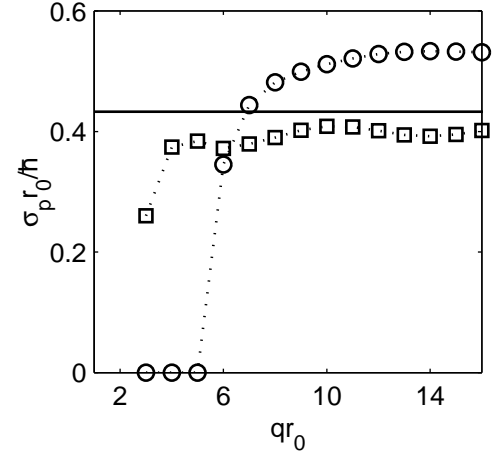} \par}

\caption{\label{FIG_momentum_width} The true momentum width
\protect\( \sigma _{p}\protect \)
of the condensate (solid line) in comparison to the estimated value from
Eqs. (\ref{EQN_quad_red_width}) (squares) and
(\ref{EQN_MIT_quad})
(circles). Other parameters as in Fig. \ref{FIG_srf_moments}.}
\end{figure}

Our approximate form $R_{QH}$ provides a more accurate means of  extracting
the Doppler width. We first assume that the finite time and density effects
are well accounted for by $R_{QH},$ which has an an rms-width  $\Delta R_{QH}$.
We then assume that the overall width is obtained by adding
this in quadrature to the Doppler width, which gives  the following  estimate
$\tilde{\sigma}_{p}$ for the momentum width
\begin{equation}
\tilde{\sigma}_{p}=\frac{m\,\sqrt{m_{\mathrm{rms}}^{2}-(\Delta R_{QH})^2
}}{q}  \label{EQN_quad_red_width}
\end{equation}
in which no fitting parameter is necessary. Results for
$\tilde{\sigma}_{p}$ are also shown in Fig.
\ref{FIG_momentum_width} and clearly give a better estimate for
the true width $\sigma _{p}$ than Eq. (\ref{EQN_MIT_quad}).

\subsection{Scattering beyond the linear regime}
\label{nlresp}

\begin{figure}
{\centering \includegraphics{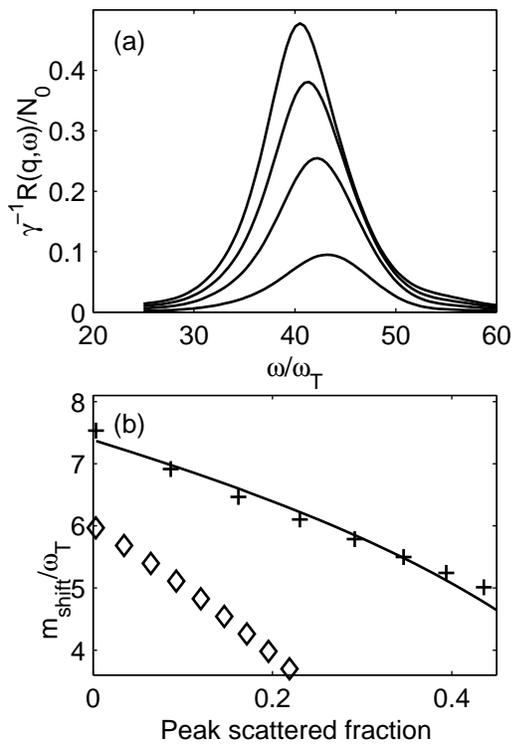} \par}

\caption{\label{FIG_finite_scatter_effect} Laser intensity effects
on the spectral response function. (a) Total scattered condensate
fraction for \protect\( q=6/r_{0}\protect \) and (lowest to
highest) \protect\( V_{p}=0.51,\, 0.85,\, 1.1,\, 1.3\, \omega
_{T}.\protect \) (b) Shift of peak response frequency \protect\(
m_{\rm {shift}}\protect \) as a function of the peak scattered
fraction for \protect\( q=3/r_{0}\protect \) (diamonds),
\protect\( 6/r_{0}\protect \) (solid) and \protect\(
9/r_{0}\protect \) (crosses). Other parameters: spherically
symmetric condensate, \protect\( N_{0}U_{0}=10^{4}w_{0},\protect
\) \protect\( q_{0}=3.8/r_{0}\protect \) and \protect\(
T_{p}=0.8/\omega _{T}\protect \). }
\end{figure}
Spectroscopic experiments require that the number of atoms scattered be
sufficient for the clear identification of momentum transfer to the system.
In \cite{Strenger99} the light intensity was chosen so that the largest
amount of condensate scattered (at the Bragg resonance) was about $20\%$.
The validity of the linear theories (e.g. section \ref{SEC_THEORY}) must be
questioned at such large fractional transfers, where excitations can no
longer be considered noninteracting (e.g. see \cite{Morgan98}). The
numerical GPE calculations do however remain valid in this regime, and can
be used to understand the changes in the spectral response function that
occur as the scattered fraction increases.

In Fig. \ref{FIG_finite_scatter_effect} we present results which
extend beyond the linear regime. In Fig.
\ref{FIG_finite_scatter_effect}(a) we plot the scattered
condensate fraction [$\gamma ^{-1}R(\mathbf{q},\omega )/N_0]$ in a
sequence of curves for increasing $V_{p}$. We have chosen to use
the
scattered condensate fraction rather than the spectral response function $R(%
\mathbf{q},\omega )$, since the numerical value of the scattered fraction
indicates whether the measurement is outside the linear regime. This is
shown clearly in this sequence of curves, which would simply be scaled by
the ratios of $V_{p}^{2}$ if they were in the linear regime. Instead, we see
that the shapes of the curves change as $V_{p}$ increases (and the scattered
fraction increases) and in particular the peak frequency shifts downwards.
In Fig. \ref{FIG_finite_scatter_effect}(b) we consider that dependence of
the mean response frequency shift ($m_{\mathrm{shift}}$) on the peak
scattered fraction of condensate (i.e. the fraction of condensate scattered
at the Bragg resonance for the curves) for three different $q$ values. A
number of features are apparent in those curves. First, the shift decreases
below the linear prediction as the scattered fraction increases. Second,
the decrease is larger for low momentum transfers ( $q<q_{0})$ than for
higher momentum transfers. Third, for large enough momentum transfer, the
shift at a given condensate fraction saturates

We note that in the MIT experiments a peak excitation fraction of $\sim 0.2$
was used \cite{Strenger99}. From Fig. \ref{FIG_finite_scatter_effect} we see
that even for the highest values of momentum transfer, (i.e. $q=6/r_{0}$ or $%
q=9/r_{0}$), the shift will be reduced by of order $15\%$ relative to the
prediction of the linear theory.

\section{Long time excitation: energy
response}\label{SEC_long_time}

\begin{figure}
{\centering
\includegraphics{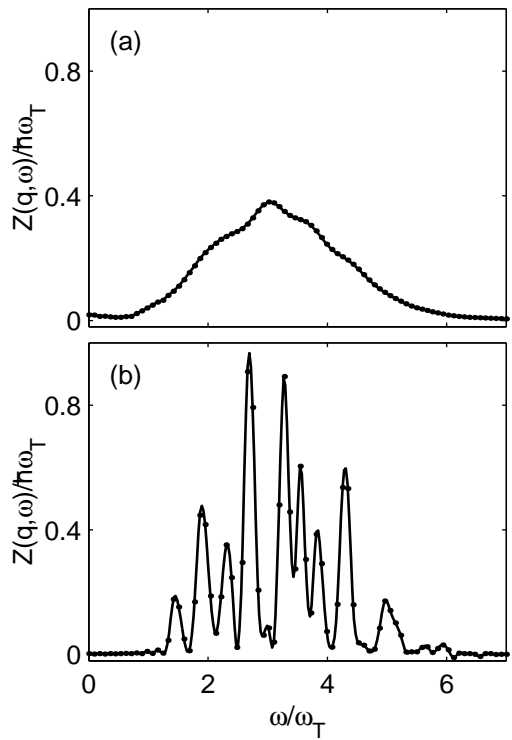} \par}

\caption{\label{FIG_long_time_excitation} Energy response function
calculated for a two dimensional condensate and a Bragg grating
with \protect\( q=1/r_{0}\protect \) and duration (a) \protect\(
T_{p}=2\pi /\omega _{T}\protect \) and (b) \protect\( T_{p}=10\pi
/\omega _{T}\protect \). In both cases the intensity is chosen so
that \protect\( \gamma^{-1} \approx 0.05N_0\protect \).}
\end{figure}

As discussed in section \ref{SECmeasurecond}, momentum is not a
conserved quantity in a trapped condensate, and is not a
convenient observable for the long time limit. Here we investigate
the energy transfer to the condensate by Bragg excitation (see
also \cite{Brunello2001}), since in the unperturbed system (i.e.
in the absence of the Bragg grating) energy is a conserved
quantity\footnote[3]{%
We ignore the effect of collisional losses from the condensate
which should be small for the time scales we consider here.}. A
possible method of measuring the energy transferred is the
technique of calorimetry, which has been applied by the Ketterle
group in a somewhat different context
\cite{WolfgangCommunication}.

We define an energy response function $Z(\mathbf{q},\omega)$
analogous to the (momentum) spectral response function by the
equation
\begin{equation}
\label{EQN_erf_defn}
Z(\mathbf{q},\omega )\equiv\gamma \, \left[ E[\Psi
(\mathbf{r},T_{p})]-E[\Psi (\mathbf{r},t=0)]\right] ,
\end{equation}
 where\begin{equation}
\label{EQN_energy_functional}
E[\Psi ]=\int d\mathbf{r}\, \left[ \Psi ^{*}\left[ -{\hbar ^{2}\over
2m}\nabla ^{2}+V_{T}(\mathbf{r})\right] \Psi +{U_{0}\over 2}|\Psi
|^{4}\right] ,
\end{equation}
is the energy functional and \( \gamma  \) is   defined in Eq.
(\ref{EQN_gamma_defn}).
Thus \( Z(\mathbf{q},\omega ) \) is the rate of energy transfer to
the condensate from a Bragg grating of frequency \( \omega  \) and
wave vector \( \mathbf{q} \). The results of section
\ref{SEC_THEORY}
for the quasiparticle evolution can be readily applied to evaluating
Eq. (\ref{EQN_erf_defn}). Using the \( T=0 \)K results of
either the vacuum expectation of the Bogoliubov Hamiltonian in Eq.
(\ref{EQN_H0_BOG}) with \( \hat{b}_{i} \) given by
(\ref{EQN_bi_from_Heisen_Eq_of_Mot}),
or the linearized Gross-Pitaevskii result
(\ref{EQN_ci_from_GP_soln})
in the energy functional (\ref{EQN_energy_functional}) gives
\begin{equation}
\label{EQN_energy_sum}
E[\Psi (\mathbf{r},t)]=E_{0}+\sum _{i}\hbar \omega _{i}\langle
\hat n_{i}(t)\rangle ,
\end{equation}
where \( E_{0}=E[\sqrt{N_{0}}\psi _{0}] \) is the initial (ground
state) energy and \( \langle \hat n_{i}(t)\rangle  \) is the Bragg induced
quasiparticle occupation (\ref{EQN_MBandGP_QP_popn}) at time
\( t \). Substituting Eq. (\ref{EQN_energy_sum}) into Eq.
(\ref{EQN_erf_defn}) gives\begin{equation}
\label{EQN_erf_ni}
Z(\mathbf{q},\omega )=\gamma \sum _{i}\hbar \omega _{i}\langle
\hat n_{i}(T_{p})\rangle .
\end{equation}

As noted above, because energy is a constant of motion of the trapped
condensate, we can take \( T_{p}\gg 2\pi /\omega _{T} \) (though
we require \( V_{p}^{2}T_{p} \) to remain small compared to unity
for our linear analysis to remain valid) and from Eq.
(\ref{EQN_pop_general_square})
(with \( \langle \hat n_{i}(0)\rangle =0 \)), we have \begin{equation}
\label{EQN_Z_interms_of_ui_and_vi}
Z(\mathbf{q},\omega )=\sum _{i}\hbar \omega _{i}\left| \int
d\mathbf{r}(u^{*}_{i}+v^{*}_{i})e^{i\mathbf{q}\cdot \mathbf{r}}|\psi
_{0}|\right| ^{2}F(\omega _{i}-\omega ,T_{p}).
\end{equation}
Thus in long duration (weak) Bragg excitation, the only contribution
to \( Z(\mathbf{q},\omega ) \) will come from quasiparticle states
with energy approximately matching \( \hbar \omega  \) (see Eq.
(\ref{EQN_ni_LONG_TIME})).

In Fig. \ref{FIG_long_time_excitation} we present results
for the energy response function of a 2D condensate for two different
durations of Bragg excitation. Fig.
\ref{FIG_long_time_excitation}(a)
shows the case of Bragg excitation applied for a single trap period
(\( T_{p}=2\pi /\omega _{T} \)). In this case the frequency spread
of the Bragg pulse is sufficiently wide that only a single broad peak
of energy absorption is discernible from the condensate i.e. the
functions
\( F(\omega -\omega _{i},T) \) appearing in summation of Eq.
(\ref{EQN_Z_interms_of_ui_and_vi})
are sufficiently broad in frequency space to allow a large number
of quasiparticles to respond. In Fig.
\ref{FIG_long_time_excitation}(b)
\( Z(\mathbf{q},\omega ) \) is shown for the case of a Bragg
pulse applied for five trap periods (\( T_{p}=10\pi /\omega _{T} \)).
Here the detailed structure of the energy response function is revealed,
with individual (possibly degenerate) quasiparticle peaks being
visible. We emphasize that current
momentum response experiments (i.e. \( R(\mathbf{q},\omega ) \)) are
essentially limited to at most a quarter trap period of excitation.

A response measurement such as shown in Fig.
\ref{FIG_long_time_excitation}(b)
reveals a wealth of information about the nature of the condensate
excitation spectrum. The frequencies (\( \omega _{i} \)) of the
quasiparticles
can be determined by the location of the resonant peaks of the
response function, and these frequency
peaks become narrower and better defined as $T_p$ increases.
We note that for a given peak corresponding to a Bragg frequency
\( \omega _{p} \), the area under the peak is equal to the matrix
element \begin{equation}
\label{EQN_area_about_QP}
\int _{\delta \omega }d\omega \, Z(\mathbf{q},\omega )=\sum _{i\,
(\omega _{i}\in \delta \omega )}\hbar \omega _{i}\left| \int
d\mathbf{r}(u^{*}_{i}+v^{*}_{i})e^{i\mathbf{q}\cdot \mathbf{r}}|\psi
_{0}|\right| ^{2},
\end{equation}
 where \( \delta \omega  \) (\( \approx [\omega _{p}-\pi
/T_{p},\omega _{p}+\pi /T_{p}] \)) is the frequency width of the
pulse  and the summation is taken over quasiparticles in the
energy range \( \delta \omega  \).


\section{Discussion}

In this paper we have given a detailed theoretical analysis of the
phenomenon of Bragg spectroscopy from a Bose-Einstein condensate. We began
by deriving analytic expressions for the evolution of the quasiparticle
operators, which contain all possible information about the system in the
linear response regime. We then demonstrated that at $T=0$K, the mean values
of the quasiparticle operators are identical to the quasiparticle amplitudes
obtained by solving the linearised Gross-Pitaevskii equation. Thus, for the
purpose of calculating the observable of the Bragg spectroscopy experiments
(the transferred momentum), a meanfield treatment is equivalent to the full
quantum treatment. Consequently, we based our detailed analysis of Bragg
spectroscopy on the meanfield equation for Bragg scattering presented in a
previous paper. The central object for the experiments is the spectral
response function for the momentum transfer, $R(\mathbf{q},\omega ).$ We
derived the relationship between $R(\mathbf{q},\omega )$ and the dynamic
structure factor and showed that, contrary to the assumptions of previous
analyses, there is no regime in which the two quantities are equivalent for
trapped condensates.

The results of our numerical simulations of Bragg spectroscopy,
which were carried out in axially symmetric three-dimensional
cases, or in two dimensions, were characterised by the behaviour
of $R(\mathbf{q},\omega)$. These full numerical solutions are
accurate for all values of $q$ that can be resolved by the
computational grid, however the computationally intensive nature
of the calculation made quantitative comparison of our theory with
the MIT experiments difficult. The analytic approximation for the
spectral response function, $R_{QH}(\mathbf{q},\omega )$, provides
a means of extending the regime of comparison to large
condensates, and systems without axial symmetry, and we showed
that it accurately represents $R(\mathbf{q},\omega )$, except in
those regimes where the momentum width is dominant ($ \sigma
_{p}q/m$ exceeds $\pi /T_{p}$ and $\mu $) or where stimulated
effects can occur ($\hbar q<\sigma _{p}$). It also provided a
means for identifying the relative importance of three broadening
and shift mechanisms (meanfield, Doppler, and finite pulse
duration). We have shown that the suppression of scattering at
small values of $q$ observed by Stamper-Kurn \emph{et al.} \cite
{Stamper99} is accounted for by the meanfield treatment, and can
be interpreted in terms of the interference of the $u$ and $v$
quasiparticle amplitudes.

A remaining point to emphasize is that our numerical calculations  allowed
us to investigate the regime of large laser intensities where the linear
response condition is invalid. We found that a significant decrease in the
shift of the spectral response function can occur due to depletion of the
initial condensate.

\begin{acknowledgments}
This research was supported by the Marsden Fund of New Zealand
under contract PVT902. PBB is grateful to Dr. T. Scott and Prof.
W. Ketterle for helpful comments.
\end{acknowledgments}

\appendix

\section*{Bogoliubov
conventions\label{SEC_APPEND_BOGOLIUBOV_Conventions}}

Several different forms of the Bogoliubov transformation have been
used in the theoretical description of inhomogeneous Bose-Einstein
condensates. These transformations typically differ in their choice
of sign between the \( u \) and \( v \) amplitudes,
and by the explicit inclusion of phase. To assist comparison
of our results to the work of others, we summarize four
different definitions,
and show how the form of the quasiparticle population result (Eq.
(\ref{EQN_MBandGP_QP_popn})) is altered by each choice. In
section \ref{SEC_BOG_TRANSFORMATION} we used the orthogonal
quasiparticle basis, since the Bogoliubov diagonalization of the
many-body
Hamiltonian must be made with excitations that are orthogonal to
the ground state. However, the quasiparticle population result, Eq.
(\ref{EQN_MBandGP_QP_popn}), is insensitive to this and so
for brevity we will use the non-orthogonal basis states.

In what follows we denote our choice of quasiparticle amplitudes by
the notation \( \{u^{(+)}_{i},v_{i}^{(+)}\} \). These are closely
related to the form used in \cite{Fetter96} by Fetter (which we denote \(
\{u^{(-)}_{i},v_{i}^{(-)}\} \)),
but differ by a minus sign in the relative phase between \( u_{i} \)
and \( v_{i} \). Both these conventions explicitly include the
condensate
phase, and the operator part of the field operator is expanded in the
form\begin{equation}
\label{EQN_decomp_uv}
\hat{\phi }=e^{iS_{0}(\mathbf{r})}\sum _{i}\left[ u^{(\pm
)}_{i}(\mathbf{r})\hat{b}_{i}(t)\pm v_{i}^{(\pm
)*}(\mathbf{r})\hat{b}_{i}^{\dagger }(t)\right] ,
\end{equation}
 where the quasiparticles modes obey the equations (at \( T=0 \)K
with \( N_{0} \) particles in the \( \psi _{0} \)
state)\begin{eqnarray}
{\cal L}u^{(\pm )}_{i}\pm N_{0}U_{0}|\psi _{0}|^{2}v_{i} & = & \hbar
\omega _{i}u^{(\pm )}_{i},\label{EQN_uv} \\
{\cal L}^{*}v^{(\pm )}_{i}\pm N_{0}U_{0}|\psi _{0}|^{2}u^{(\pm )}_{i}
& = & -\hbar \omega _{i}v^{(\pm )}_{i},\label{EQN_uv2}
\end{eqnarray}
and \( {\cal L} \) is as defined in Eq. (\ref{EQN_calL}).
The basis choice of Eqs. (\ref{EQN_uv}) and
(\ref{EQN_uv2})
leads to the vacuum expectation value Eq.
(\ref{EQN_bi_expec_on_vac})
having the form\begin{eqnarray}
\langle \hat{b}_{i}(t)\rangle&=&-i\sqrt{N_{0}}\int _{0}^{t}dt^{\prime
}\, V(t^{\prime })e^{i\omega _{i}t^{\prime }}\int d\mathbf{r}\,
(u^{(\pm )*}_{i}\pm v^{(\pm )*}_{i})\nonumber\\
&&\times\cos (\mathbf{q}\cdot
\mathbf{r}-\omega t)|\psi _{0}|.
\end{eqnarray}
The most common form of the Bogoliubov transformation (e.g. see
\cite{Griffin96, Morgan98}),
uses the following form for the expansion of the operator part of
the field \begin{equation}
\label{EQN_decomp_uvbar}
\hat{\phi }=\sum _{i}\left[ \bar{u}^{(\pm
)}_{i}(\mathbf{r})\hat{b}_{i}(t)\pm \bar{v}_{i}^{(\pm
)*}(\mathbf{r})\hat{b}_{i}^{\dagger }(t)\right] ,
\end{equation}
where the \( \pm  \) indices indicate the relative choice of sign
between the \( u \) and \( v \) terms in Eq.
(\ref{EQN_decomp_uvbar}).
These quasiparticles modes obey the equations (at \( T=0 \)K with
\( N_{0} \) particles in the \( \psi _{0} \) state)\begin{eqnarray}
L\bar{u}^{(\pm )}_{i}\pm N_{0}U_{0}|\psi _{0}|^{2}\bar{v}^{(\pm
)}_{i} & = & \hbar \omega _{i}\bar{u}^{(\pm
)}_{i},\label{EQNubar} \\
L\bar{v}^{(\pm )}_{i}\pm N_{0}U_{0}|\psi _{0}|^{2}\bar{u}^{(\pm
)}_{i} & = & -\hbar \omega _{i}\bar{v}^{(\pm )}_{i},
\end{eqnarray}
where \( L \) is defined as\begin{equation}
\label{EQN_L}
L=\left[ -{\hbar ^{2}\over 2m}\nabla ^{2}+V_{T}(\mathbf{r})-\hbar \mu
+2N_{0}U_{0}|\psi _{0}|^{2}\right] .
\end{equation}
In this case the mean value of \( \hat{b}_{i} \) is
\begin{eqnarray}
\label{EQN_bi_uvbar}
\langle \hat{b}_{i}(t)\rangle &=&-i\sqrt{N_{0}}\int _{0}^{t}dt^{\prime
}\, V(t^{\prime })e^{i\omega _{i}t^{\prime }}
\BR\times\int d\mathbf{r}\,
(\bar{u}^{(\pm )*}_{i}\psi _{0}\pm \bar{v}^{(\pm )*}_{i}\psi
_{0}^{*})\cos (\mathbf{q}\cdot \mathbf{r}-\omega t),\BR
\end{eqnarray}
 (see Eq. (\ref{EQN_bi_expec_on_vac})). When \( \psi _{0} \)
is a ground state (i.e. has constant phase), \( \psi _{0} \) can be
taken as real in Eq. (\ref{EQN_bi_uvbar}),
i.e.
\begin{eqnarray}
\label{EQN_bi_uvbar2} \langle \hat{b}_{i}(t)\rangle &=&
-i\sqrt{N_{0}}\int _{0}^{t}dt^{\prime }\, V(t^{\prime })e^{i\omega
_{i}t^{\prime }} \BR\times \int d\mathbf{r}\, (\bar{u}^{(\pm
)*}_{i}\pm \bar{v}^{(\pm )*}_{i})\cos (\mathbf{q}\cdot
\mathbf{r}-\omega t)\psi _{0}.\BR
\end{eqnarray}

\bibliography{Paper5}
\end{document}